\documentclass[a4paper, 12pt]{article}
\usepackage{amsfonts,amsmath,amsthm}
\usepackage{graphicx,authblk,hyperref}
\usepackage{microtype}
\usepackage{lipsum}
\usepackage[section]{placeins}
\usepackage{afterpage}
\usepackage{pifont}
\usepackage{bm}
\usepackage{multirow}
\usepackage{pdfpages}

\newtheorem{thm}{Theorem}
\newtheorem{lem}[thm]{Lemma}

\newtheorem{prt}[thm]{Property}
\newtheorem{algo}{Algorithm}

\def\NN{{\mathbb N}}

\begin{document}

\title{The cross-sectional distribution of portfolio returns and applications}
\author[1]{Ludovic Cal\`es}
\author[2,3,4]{Apostolos Chalkis}
\author[3,2]{Ioannis Z.~Emiris}

\affil[1]{European Commission, Joint Research Centre, Ispra, Italy}
\affil[2]{Department of Informatics \& Telecommunications\newline National \& Kapodistrian University of Athens, Greece}
\affil[3]{ATHENA Research \& Innovation Center, Greece}
\affil[4]{GeomScale org}
\date{}
\maketitle

\begin{abstract}
This paper aims to develop new mathematical and computational tools for modeling the distribution of portfolio returns across portfolios. We establish relevant mathematical formulas and propose efficient algorithms, drawing upon powerful techniques in computational geometry and the literature on splines, to compute the probability density function, the cumulative distribution function, and the $k$-th moment of the probability function. Our algorithmic tools and implementations efficiently handle portfolios with $10\, 000$ assets, and compute moments of order $k\leq 40$ in a few seconds, thus handling real-life scenarios. We focus on the long-only strategy which is the most common type of investment, i.e.\ on portfolios whose weights are non-negative and sum up to 1; our approach is readily generalizable. Thus, we leverage a geometric representation of the stock market, where the investment set defines a simplex polytope. 
The cumulative distribution function corresponds to a portfolio score capturing the percentage of portfolios yielding a return not exceeding a given value. We introduce closed-form analytic formulas for the first 4 moments of the cross-sectional returns distribution, as well as a novel algorithm to compute all higher moments. We show that the first 4 moments are a direct mapping of the asset returns' moments. All of our algorithms and solutions are fully general and include the special case of equal asset returns, which was sometimes excluded in previous works. 
Finally, we apply our portfolio score in the design of new performance measures and asset management. We found our score-based optimal portfolios less concentrated than the mean-variance portfolio and much less risky in terms of ranking.

\bigskip
\noindent Keywords:
Portfolio return distribution, Moments, Portfolio score, Finance, Geometry, B-spline, Algorithms, Computational experiment
\end{abstract}


\section{Introduction}

\begin{figure}[h!] \centering
    \includegraphics[width=0.65\textwidth]{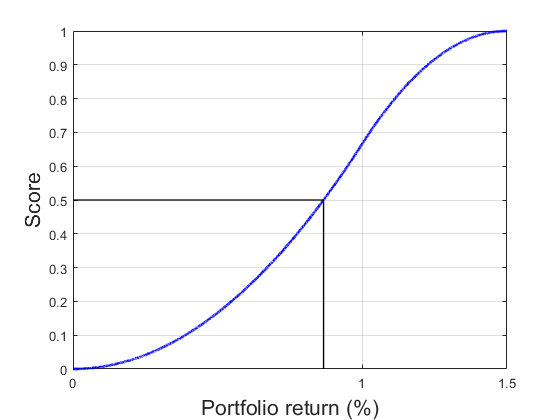}
\caption{Score of a long-only portfolio in a market of 3 assets whose returns are $0\%$, $1\%$, and $1.5\%$.\label{fig:IllustrationScore}}
\end{figure}

The study of the probability distribution of portfolio returns, across portfolios and for given asset returns, has attracted less attention than it deserves in the finance literature and probability theory. 
However, it is a natural tool to understand the relative performance of portfolios, as well as market dynamics in general.
Indeed, consider an investment set corresponding to the long only strategy, defined such that the portfolio weights are non-negative and sum up to~$1$.
Let us define the -cross sectional- {\em score} of a portfolio as the percentage of portfolios, within the investment set, that this portfolio outperforms -in terms of return- in a given time period.
This score measures its performance relative to the other portfolios. 
Thus, we consider it as the value of the cumulative distribution function (CDF) for a given value of return.
We use the term {\em cross section} to underline that the score is defined for a given period of time and a set of portfolios.
For instance, in a market of 3 assets whose returns are $0\%$, $1\%$ and $1.5\%$, the score of a portfolio as a function of its return is presented in Figure \ref{fig:IllustrationScore}.
Here, a portfolio whose return is $0.866\%$ outperforms $50\%$ of the portfolios.
Thus, it allows for inter-temporal comparison of portfolio performance.

{\bf Literature review and motivation.} 
The resilience of the asset management industry during the past few decades has highlighted the analysis of portfolio allocation performance as an important aspect of modern finance.
Research in this area is axed on Sharpe-like ratios proposed in the 1960’s \cite{Jensen67,Sharpe66,Treynor15}.
Managers are typically ranked according to these ratios, and those achieving the highest and steadiest returns receive the top scores.
The major drawback of these techniques is the identification of benchmark portfolios, while the formation of such portfolios remains controversial.
Moreover, existing approaches suffer from significant estimation errors, see e.g.~\cite{Lo05}, which prevent any performance comparison to be significant.

The portfolio score considered in this paper has been introduced in \cite{P05} and has been used in related studies by the same author.
In~\cite{P05,PST04,HHPS2002}, the relative performance of value-weighted indices with respect to long-only portfolios is assessed in the Dutch, Spanish and German markets; they considered the MSCI Netherlands 24, IBEX~35, and DAX~30 components, respectively.
It leads the authors to question the representativeness of these indices.
In~\cite{HP2005,HP16}, the dispersion of the cross-sectional portfolio returns is used to assess the performance of asset managers whose mandate implies tracking error volatility constraints.
This score has also been proposed independently in \cite{BCG11}, and in \cite{BH11}.
In~\cite{BCG11}, this score is used as a portfolio performance measure whose informativeness is shown to be higher than the Sharpe and the Sortino ratios in an asset allocation exercise.
In \cite{BH11}, a simplified score assigning a reward ranging from -2 to 2 according to the quintile of the score is used to assess the performance of the momentum strategy.
This investment strategy is shown to not be outperforming an uninformed naive investor.
The score is also shown to be invariant under affine transformation which means that it is not affected by a shift in the asset returns nor by a change in the returns' scale.
It makes the score robust against shocks hitting equally all asset returns, such as risk-free shocks in the classical APT model, and volatility jumps.
Recently, in \cite{CCEF2018}, the score is used to study the time-varying dependency of portfolios' return and volatility, and relates this dependency to periods of financial turmoils.

Thus, the CDF of the portfolio returns across portfolios is important for various applications in asset management, portfolio performance measurement, and for studying financial stability.

We also believe that the moments of this distribution would find applications in the literature on return dispersion, see e.g.\ \cite{yu2007alpha}, ~\cite{stivers2010cross}, ~\cite{gorman2010cross}, ~\cite{bhootra2011momentum} as well as ~\cite{verousis2018cross};
on noise trading, see e.g.~\cite{DSSW89};
and on event studies, see e.g.~\cite{BOEHMER1991253}.
This wide range of financial applications emphasizes the importance of a formal and rigorous statistical characterization of the portfolio returns' distribution, which we provide here.

{\bf Current computational techniques.}
In \cite[Thm~4.2.2]{P05} a geometry-based closed form expression of the score is proposed.
It consists of representing the long-only investment set as a simplex, just as in this paper.
The score is then the volume of the intersection of the simplex and a linear half-space.
It is computed by decomposing this intersection in smaller simplices.
Following this approach, the score of a portfolio whose return is $R$, in a market of $n$ assets whose returns are $(R_i)_{i=1}^n$, can be computed as follows:
\begin{equation}\label{eq:Pouchkarev}
\mbox{Score} (R) = \sum\limits_{R_k\leq R} \left( (R-R_k)^{n-1} \prod\limits_{i=1, i\neq k}^n \frac{1}{R_i-R_k} \right) .
\end{equation}
However, this computation is not valid when some asset returns are equal and it presents floating point errors limiting its use to around 20 assets.   
As a consequence, in \cite{P05} and in related studies, the score is estimated by a quasi-Monte Carlo sampling of the portfolios; one may refer to~\cite{RbMel98} for uniform sampling methods over a simplex of general dimension.

In~\cite[Thm~A2]{BH11}, the same approach is considered for illustration purposes only, and the authors also rely on portfolio sampling for their application.
They again assume unequal returns.
However, note that the formula originally proposed in \cite{BH11} contains a couple of mistakes: the sum was over the number of assets whose returns are lower than the return of the portfolio considered, and the term within parentheses in the numerator should have been the opposite.

In~\cite{BCG11}, the set of portfolios considered is the specific set of long/short equally weighted zero-dollar portfolios. 
The estimation of the score relies on combinatorics and order statistics, and it is computationally limited to around 20 assets.
Finally, in \cite{CCEF2018}, the score is also computed as the volume of the intersection of the simplex and a linear half-space.
The authors noticed that an algorithm by~\cite{Varsi} can be used to compute this volume exactly and efficiently for any number of assets, even when some asset returns are equal.

{\bf Contribution.}
In this paper, we study the distribution of portfolios' returns, by providing novel mathematical formulas and efficient computations for the CDF, the Probability Density Function (PDF) and the $k$-th moment of the probability function. We consider, for the first time, the aforementioned portfolio score as the value of the CDF for a given value of return. 
In all cases, we allow for the special scenario of equal asset returns, a case that had not been always treated in the past. We consider the long-only strategy, although the same methodology can be used to study the portfolio return distribution for any other set of investments.

We compute exactly the returns' CDF and PDF at any point.
For the former, we show (Section~\ref{sec:CDF}) that Varsi's algorithm \cite{Varsi} settles the problem for real-life situations since it is sufficiently accurate and fast even for large stock markets. In particular, when the number of assets $n=1,000$ we compute the CDF at any point in $\approx 3\cdot 10^{-3}$~sec (see Table~\ref{tab:CompTime_CDF}).
Furthermore, we demonstrate that Varsi's algorithm largely outperforms an alternative closed-form formula from \cite{Lasserre2015VolumeOS}; the latter was recently rectified in \cite{C2019} to include the case of equal asset returns, as opposed to \cite{P05,BH11} which assume distinct returns.
In addition, Lasserre's closed-form formula is inaccurate for $n\geq 20$, hence the approach of this paper is the method of choice for realistic applications.

For the computation of PDF we provide two novel methods for exact computation at any point. Our first approach is based on the geometric interpretation of univariate B-splines by \cite{CS66}, and it leverages the de Boor-Cox recursive formula, see \cite{dB72} and \cite{C72}.
Our second approach is a direct derivation, obtained via the closed-form formula in~\cite{Lasserre2015VolumeOS}.
Moreover, to overcome possible numerical inaccuracies of both previous methods while $n$ increases --due to floating point arithmetic-- we employ Taylor's expansion and Varsi's algorithm for approximate computation of PDF, at any point, within arbitrarily small error. 
Indeed, the latter approach is shown to scale up to $n = 10,000$.

In order to compute the moments of portfolio returns distribution, we rely on Theorem~\ref{thm_Lass} (Section~\ref{Lasserre} of this paper). The theorem had been demonstrated in~\cite{LA2001} to express the $k$-th order moment as a nested sum.
This is too costly to be calculated straightforwardly, in terms of complexity. Thus, we derive simpler analytic formulas, which require a number of operations proportional to $n$, for computing the first four moments.
For higher orders, we employ the identity in \cite{MacDonald1995} to obtain a novel expression which requires computing the partitions of integer $k$, for the $k$-th moment.
Thanks to a result in~\cite{ZS1994} for computing partitions, our algorithm obtains the moment when $k=40$ and $n=10,000$ in a few milliseconds (see Table~\ref{tab:CompTimeMoments}). We also prove direct mappings between the first four moments of portfolio returns distribution and the respective moments of the asset returns distribution, while we make several useful observations on these mappings.

\begin{figure}[t]
\centering
\begin{minipage}[h]{0.46\textwidth}
\includegraphics[width=\linewidth]{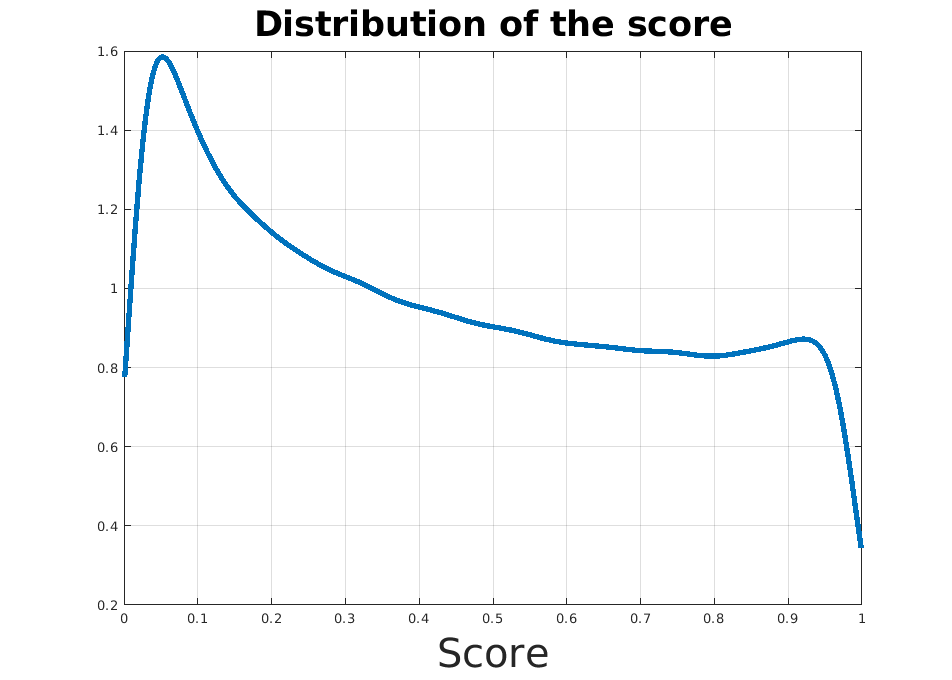}
\end{minipage}
\hspace{0.6cm}
\begin{minipage}[h]{0.46\textwidth}
\includegraphics[width=\linewidth]{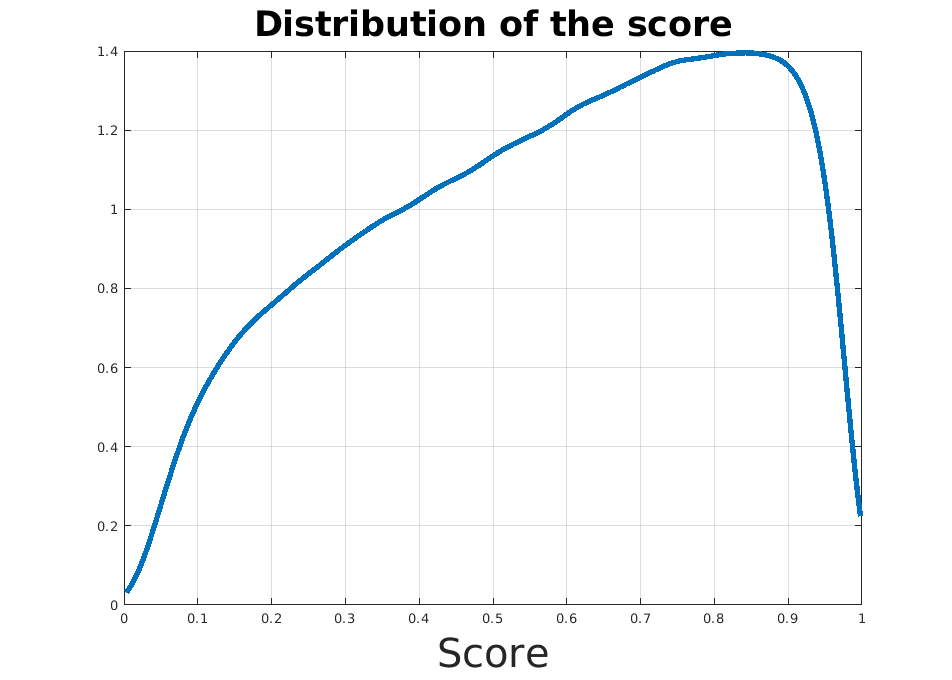}
\end{minipage}
\caption{The PDF of the score of two portfolios. The portfolio on the left-hand side shows a mass to the left of the distribution, indicating it is often outperformed by the portfolios of the investment set. 
However, the one on the right appears to frequently outperform the other portfolios. \label{fig:example_dist_scores}}
\end{figure}

Last but not least, we show the relevance of the score in asset allocation.
We estimate the distribution of a portfolio' score as in Figure~\ref{fig:example_dist_scores}.
With this distribution, we can quantify the risk for a portfolio to be among the worst performers and its opportunity to be among the best ones.
We show that the score distribution offers a new way to distinguish portfolios, even when portfolios have identical distributions.
We use the score distribution to propose new performance measures which are later used in a portfolio optimisation problem.
The score-based optimal portfolios are found to be less concentrated than the mean-variance (MV) portfolio and much less risky in terms of ranking.  
This is confirmed in a pseudo-real time asset allocation exercise where the score-based optimal portfolios also achieve a performance very close to the equally-weighted risk contribution (ERC) portfolio.  

\textbf{Paper structure.}
In Section~\ref{sec:Geometry}, we formalize the representation of the investment set as a unit simplex.
The portfolios considered are then uniformly distributed over this simplex, allowing us to integrate over it, later on. 
Section~\ref{sec:CDF} presents efficient computations of CDF, and discusses some important properties of the score.
Section~\ref{sec:PDF} offers novel methods for exact and approximate computation of the PDF.
Section~\ref{sec:Moments} presents our closed-form formulas for the first four moments, while it also introduces new algorithms to compute the $k$-th order moment of the portfolio returns distribution, for any $k\ge 1$. 
Section~\ref{sec:dist_score} discusses the usefulness of the score in asset allocation through detailed examples and presents new portfolio performance measures.
We conclude with an overall discussion and open questions.
The Annexes contain some mathematical proofs and complementary materials for Section~\ref{sec:dist_score}.

\section{Geometric representation of the set of portfolios}\label{sec:Geometry}

In this section we formalize the geometric representation of sets of portfolios with an arbitrary number of assets.

Let us consider a portfolio $x$ investing in $n$ assets, whose weights are $x=(x_1, \dots, x_n)$. 
The portfolios in which a long-only asset manager can invest are subject to $\sum\limits_{i=1}^{n} x_i = 1$ and $x_i\geq 0, \forall i$. 
Thus, the set of portfolios available to this asset manager is the unit $(n-1)-$dimensional simplex, denoted by $\Delta^{n-1}$ and defined as
\begin{equation}\label{eq:1}
\Delta^{n-1} = \left\{ \sum_{i=1}^{n} x_i v_i \left| (x_1, \dots, x_n) \in \mathbb{R}^{n}, \sum_{i=1}^{n} x_i=1, \mbox{ and } x_i \ge 0, \forall i \in \{1, \dots, n\} \right.\right\} ,
\end{equation}
where $v_1, \dots, v_n \in \mathbb{R}^{n-1}$ are a set of $n$ affinely independent points in a Euclidean space of dimension $n-1$. The vertices $(v_i)_{i=1,\dots,n}$ represent the $n$ portfolios made of a single asset and the simplex is the convex hull of these vertices. \\

\noindent For instance, we can define $v_1, \dots, v_n$ such that:

\begin{enumerate}
	\item the center of the simplex is set to the origin,
	\item the distances of the simplex vertices to its center are equal,
	\item the angle subtended by any two vertices through its center is $\arccos(\frac{-1}{n-1})$.
\end{enumerate}

\noindent 
The weights $(x_i)_{i=1,\dots,n}$ of portfolio $x$ are called its barycentric coordinates, 
whereas the corresponding coordinates in $\mathbb{R}^{n-1}$ are called its Cartesian coordinates, and are denoted by $\breve{x}=(\breve{x}_1, \dots, \breve{x}_{n-1})$. 
We use the Cartesian coordinates in Section~\ref{sec:Moments} to compute the moments of the portfolios' returns distribution.


\section{The CDF of the portfolios' returns distribution}\label{sec:CDF}
In this section we focus on the exact computation of the cumulative distribution function (CDF) of the portfolio returns, given the asset returns.
Let us consider the set of long-only portfolios providing a return lower than a given return $R^*$ over a period of time for which the asset returns were $\mathbf{R}=\left(R_1, \dots, R_{n}\right)$.
It corresponds to a linear half-space defined as 
\begin{equation}
H(R^*) = \left\{ \left( x_1, \dots, x_n\right) \in \mathbb{R}^{n}\, |\, \sum\limits_{i=1}^n R_i x_i \leq R^* \right\}.
\label{eq:Hyperplane}
\end{equation}
Denoting by $V(A)$ the Euclidean volume of a geometric object $A$, the allocation score of a portfolio providing a return $R^*$ can be obtained by computing the ratio of the volume of the intersection of the simplex with this half-space over the volume of the simplex, i.e.
\begin{equation}
S(R^*) = \frac{ V(H(R^*) \bigcap \Delta^{n-1})}{V(\Delta^{n-1})} .
\end{equation}

\begin{figure*}[htbp!] \centering \begin{tabular}{cc}
\includegraphics[width=0.45\textwidth]{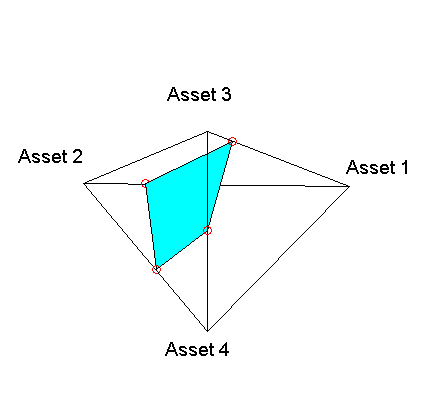}
	&
\includegraphics[width=0.45\textwidth]{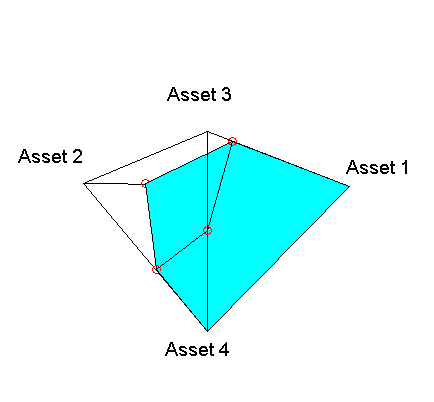}
\end{tabular}
\caption{(left) Surface (flat) of portfolios providing a given return. (right) Volume (polyhedron) of portfolios outperformed by this return. \label{tab:IllustScoreVol}}
\end{figure*}

We illustrate such a volume in Figure \ref{tab:IllustScoreVol}.
Consider a market of 4 assets whose returns are observed, and a portfolio providing a given return $R^*$. 
The pyramid is the simplex representing the set of long-only portfolios. 
The surface highlighted in the left figure represents the set of portfolios returning $R^*$.
The volume highlighted in the right figure represents the set of portfolios providing a return lower or equal to $R^*$.

\subsection{Varsi's algorithm}

As noticed in~\cite{CCEF2018}, there exists an exact, iterative formula for the volume defined by intersecting a simplex with a half-space. 
It is provided in Algorithm \ref{algo_Varsi}.
A geometric proof is given in \cite{Varsi}, by subdividing the polytope into pyramids and, recursively, to simplices.
For a comparison between alternative proofs and algorithms, the reader may refer to~\cite{CCEF2018} and the references thereof.

\begin{algo}\label{algo_Varsi}

Let $H = \left\{ \left( \omega_1, \dots, \omega_n\right) | \sum_{i=1}^n R_i \omega_i \leq R^* \right\}$ be a linear half-space.
\begin{enumerate}
	\item Compute $u_i = R_i - R^*$, $i= 1,2, \dots , n$.
	\item Label the non-negative $u_j$ as $Y_1, \dots , Y_K$, and the negative ones as $X_1, \dots ,X_J$.
	\item Initialize $A_0 = 1$, $A_1 = A_2 = \dots = A_K = 0$.
	\item For $j = 1, 2, \dots , J$, repeat:
 $A_k \leftarrow \frac{Y_k A_k - X_j A_{k-1}}{Y_k - X_j}$, for $k = 1, 2, \dots , K$.
\end{enumerate}
\end{algo}

\noindent 
Then, at the last iteration $j=J$, it holds that
$$
A_K = \frac{ V(H(R^*) \bigcap \Delta_{n-1})}{V(\Delta_{n-1})}.
$$ 
Notice that $K+J=n$.

This algorithm requires $O(n^2)$ operations, and thus can be computed very quickly. 
As an illustration, we compute the score of a portfolio, whose return is $R^*=0$, in markets of $100$, $1000$ and $10,000$ assets whose returns are randomly drawn from a standard normal distribution.
The computation is repeated 1000 times. We report in Table \ref{tab:CompTime_CDF} the average computation time and its standard deviation.

\begin{table}[h!]\centering
\begin{tabular}{|l|ccc|}
\hline
Number of Assets 	& 100 & 1000 & 10,000 \\
\hline\hline
Mean computation time	& 5.89e-5	& 3.63e-3 & 0.4734\\
Standard deviation		& 1.99e-4	& 1.79e-4 & 0.0416\\
\hline
\end{tabular}
\caption{Mean computation time in seconds and standard deviation of computing the CDF at a point for markets of $100$, $1000$ and $10,000$ assets. The computations were performed using Matlab\copyright\ on a bi-xeon E2620~v3 under Windows\copyright.\label{tab:CompTime_CDF}}
\end{table}

\begin{figure*}[htbp]
	\centering
	\includegraphics[width=0.65\textwidth]{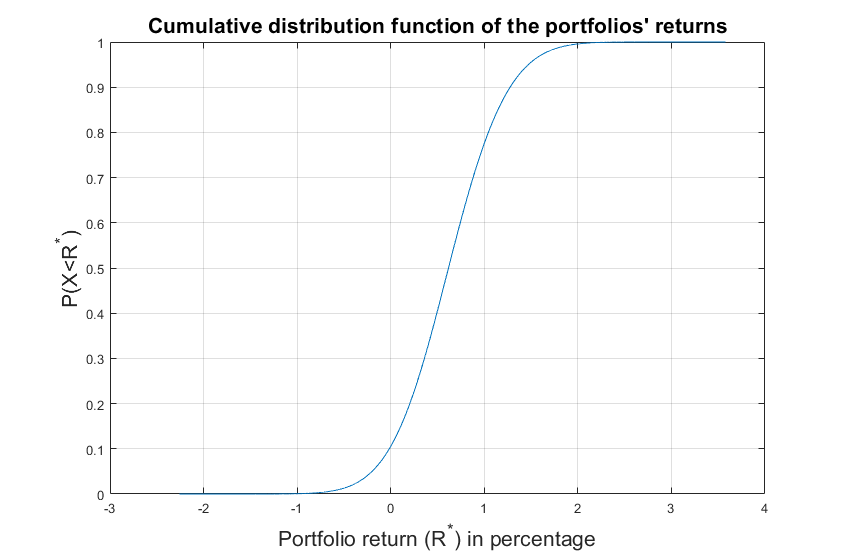}	
\caption{Illustration of the CDF of the portfolios returns, when the asset returns are given as in Table~\ref{tab:Illustration_Score}.\label{tab:IllustScore1}}
\end{figure*}

To illustrate the CDF obtained using Algorithm~\ref{algo_Varsi}, let us consider a market of 10 assets whose returns are as in Table~\ref{tab:Illustration_Score} and let $R^*$ denote a portfolio return. 
The CDF of the portfolios returns for any given return is reported in Figure~\ref{tab:IllustScore1}.
With these asset returns, $10\%$ of the portfolios have a negative return and a bit more than $20\%$ of the portfolios have a return greater than $1\%$.

\begin{table}[ht!]
\centering
\begin{tabular}{|ccccc|}
\hline
$R_1$ & $R_2$ & $R_3$ & $R_4$ & $R_5$ \\
$0.5377\%$ & $1.8339\%$ & $-2.2588\%$ & $0.8622\%$ & $0.3188\%$\\
\hline
$R_6$ & $R_7$ & $R_8$ & $R_9$ & $R_{10}$\\ 
$-1.3077\%$ & $-0.4336\%$ & $0.3426\%$ & $3.5784\%$ & $2.7694\%$\\
\hline
\end{tabular}
\caption{Some asset returns.\label{tab:Illustration_Score}}
\end{table}

\subsection{Closed form expression}\label{sec_CF_CDF}

In~\cite{Lasserre2015VolumeOS}, a closed form formula is proposed to compute this volume taking into account the case of equal asset returns.
However, it missed some extra terms and has been corrected in \cite{C2019}. 
We report it here, adapted to our notation: Let $\bm{R}=(R_i)_{i=1}^n$ be the asset returns, $(S_i)_{i=1}^d$ the $d$ distinct returns, where $d\le n$, and $(m_i)_{i=1}^d$ their multiplicities (i.e.\ number of occurrences).
We denote by $(J_i)_{i=1}^d$ the subsets of indices in $\{1, \dots, n\}$ associated to each $S_i$ and, 
$$
\mbox{for } j=1,\dots, d, \mbox{ we let } \bm{b}_{j} = \left(\frac{1}{R_i - S_j}\right)_{i \in \{1, \dots, n\} \setminus J_j}.
$$
Among the distinct returns we distinguish between those whose multiplicities are $1$, and those whose multiplicities are greater than $1$.
The indices of the first group form a set denoted by $I$, while those of the second group form set $K$, where $S=I \cup K$.
Finally, $(x)_+$ stands for $\max\{0,x\}$.
The CDF computed in $R^*$ is given by:
%
\begin{equation} \begin{array}{rl}
S(R^*, \bm{R}) =& \sum\limits_{i \in I} \frac{(R^*-S_i)_+^{n-1}}{\prod\limits_{j=1, j \neq i}^n (S_j - S_i)} +\\
    & + \sum\limits_{i \in K} \left(\sum\limits_{j=0}^{m_i-1} (-1)^{j+m_i+1} \binom{n-1}{j} \frac{(R^*- S_i)_+^{n-j-1}}{\prod\limits_{k \in S \setminus \{S_i\}} (S_k -S_i)} \Phi_{m_i-1-j}(\bm{b}_{i}) \right) , \\
\end{array}
\label{eq:Lasserre_CDF}
\end{equation}

with 
\[
\Phi_k(x) = \sum\limits_{i_1=1}^n \sum\limits_{i_2=1}^{i_1} \dots \sum\limits_{i_k=1}^{i_{k-1}} x_{i_1} \cdots x_{i_k} , \bm{x} \in \mathbb{R}^n .
\]

This formula is similar to the one proposed in \cite[Thm~4.2.2]{P05} and \cite[Thm~A2]{BH11}, but with extra terms correcting for the equal asset returns.
Unfortunately, when it comes to calculations, the formula becomes numerically unstable for $n\geq20$ at the usual machine precision.

In order to report on runtimes, we consider markets of $10$ and $20$ assets whose returns $\mathbf{R}$ are randomly drawn from a standard normal distribution.
Table~\ref{tab:CompTimeCF_CDF} provides the average computation time and its standard deviation for computing the PDF of the portfolio returns at the cross-sectional average of the asset returns, i.e.\ $R^* = \bar{R}$.
The computation is repeated 1000 times.

\begin{table}[h!]\centering
\begin{tabular}{|l|cc|}\hline
Number of Assets 		& 10 & 20 \\
\hline\hline
Mean computation time	& 3.34e-4	& 5.92e-4 \\
Standard deviation		& 3.31e-5	& 4.62e-5\\
\hline
\end{tabular}
\caption{Mean computation time in seconds and standard deviation of computing the PDF at a point for markets of $10$ and $20$ assets. The computations were performed using Matlab\copyright \ on a bi-xeon E2620 v3 under Windows\copyright.}
\label{tab:CompTimeCF_CDF}
\end{table}

\subsection{Properties of the score}

As noticed in~\cite{BH11}, the score is invariant under affine transformation of the asset returns. 
Specifically, consider a market of $n$ assets providing the returns $\bm{R}=(R_i)_{i=1}^n$. 
We are interested in the score of a portfolio $\bm{x} =(x_i)_{i=1}^n$ providing a return $R^*=\bm{x}' \bm{R}$, where $\bm{x}'$ stands for the transpose vector.
As explained before, this score is the volume of simplex $\Delta^{n-1}$ intersected with half-space $H(R^*)$ as in Equation~(\ref{eq:Hyperplane}).
So, it is 
\begin{equation}
S(R^* | \bm{R}) = \left\{ \bm{x} \in \Delta^{n-1} | \sum\limits_{i=1}^n R_i x_i \leq R^* \right\} .
\end{equation}

\begin{prt}
The score is invariant under affine transformations of the asset returns such that $ \bm{R} \rightarrow \sigma \bm{R} + \alpha $, with $\alpha \in \mathbb{R}$ and $\sigma \in \mathbb{R}^+$.
\end{prt}

\noindent \textit{Proof}.
\begin{align*}
S(\sigma R^*+\alpha\, |\, \sigma \bm{R} + \alpha)	
        & = \left\{\bm{x} \in \Delta^{n-1}\, | \sum\limits_{i=1}^n (\sigma R_i + \alpha) x_i \leq \sigma R^* + \alpha \right\} \\
	& = \left\{\bm{x} \in \Delta^{n-1}\, |\, \sigma \sum\limits_{i=1}^n R_i x_i + \alpha \leq \sigma R^* + \alpha \right\} \\
	& = \left\{\bm{x} \in \Delta^{n-1}\, |\, \sigma \sum\limits_{i=1}^n R_i x_i \leq \sigma R^*\right\} \\
	& = S(R^* | \bm{R} ) . \\
\end{align*}

The main implication of this property is that the asset returns can be standardized cross-sectionally without affecting the score.
It is interesting to note that such a transformation is common in financial event studies, since the seminal work of~\cite{BOEHMER1991253}.
This approach has the advantages of being robust to event-induced heteroskedasticity and of not requiring data from a pre-event estimation period.

\section{The PDF of the portfolios' returns distribution}\label{sec:PDF}

In this section, we focus on the exact computation of the Probability Density function (PDF) of the portfolios' returns distribution.

\subsection{Geometric interpretation of B-splines}

\begin{figure*}[h!]
	\centering
	\includegraphics[width=0.4\textwidth]{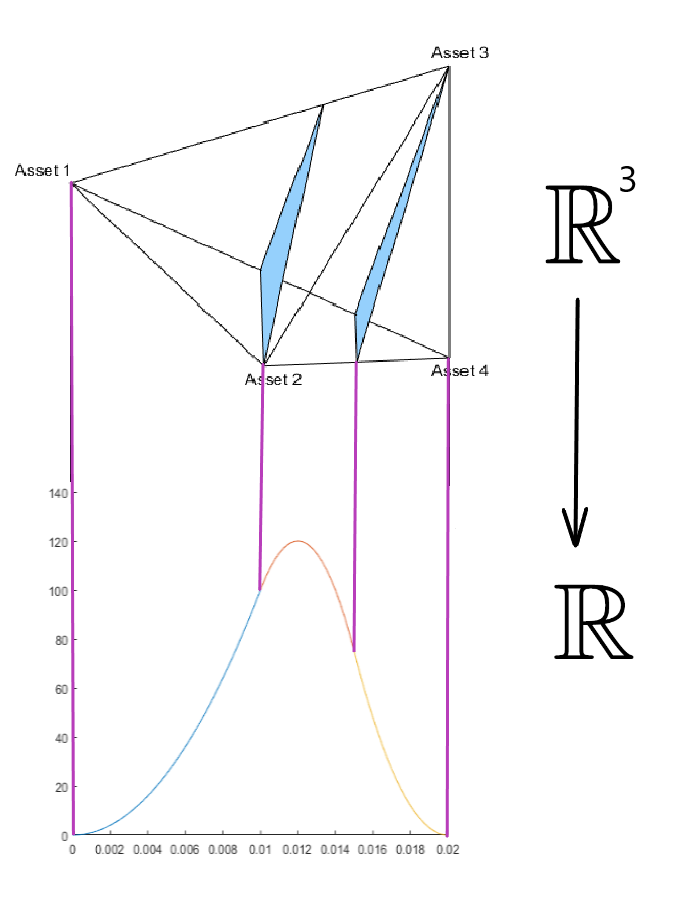}	
	\caption{Geometric interpretation of the PDF of the portfolio returns: a univariate B-spline, as the orthogonal projection of the volumetric slices of a unit simplex on $\mathbb{R}$. 
	In this example, we have 4 assets whose returns are $0\%$, $1\%$, $1.5\%$, and $2\%$. \label{tab:BSpline}}
\end{figure*}

In \cite{CS66}, a seminal paper on splines, Theorem~2 shows that the univariate B-spline resulting from the orthogonal projection of the volumetric slices of a unit simplex on $\mathbb{R}$ can be interpreted as 
the PDF of these slices' volume. 
For instance, in Figure \ref{tab:BSpline}, we have the projections of the areas of the intersection of planes with the 3-d simplex on the real line.
In our case, the simplex is the set of portfolios, while the planes are the equi-return portfolios.
The resulting univariate B-spline is the PDF of the portfolio returns.
For any value, it can be computed using the de Boor-Cox recursion formula, see \cite{dB72} and \cite{C72}, as shown in Algorithm \ref{algo_CoxDeBoor}.

\begin{algo}\label{algo_CoxDeBoor}
Let $\mathbf{R}=(R_i)_{i=1}^n$ be the returns of the $n$ assets, ordered such that $R_i \leq R_j$, for $i<j$.
To evaluate the PDF at $x$, we set the initial iterator of the spline order $m$ to $0$, the B-spline order $k$ to $n-1$ and the knot vector to $\mathbf{R}$.
Next, we execute $y= $ bspline\_pdf $(m=0, k=n-1, \mathbf{R},x)$, where function bspline\_pdf~$(\cdot)$ is specified below. 
The outcome is then normalized such that $y \leftarrow \frac{k}{R_n-R_1} y$.\\

\noindent function $y =$ bspline\_pdf $(m,k,\mathbf{R},x)$

\begin{enumerate}
	\item $y = 0$
	\item if $k>1$ then
		\begin{enumerate}
			\item $b =$ bspline\_pdf $(m,k-1,\mathbf{R},x)$;
			\item if $R_{m+k} \neq R_{m+1}$, then $y \leftarrow y + b\left(\frac{x - R_{m+1}}{R_{m+k} - R_{m+1}}\right)$
			\item $b =$ bspline\_pdf $(m+1,k-1,\mathbf{R},x)$;
			\item if $R_{m+k+1} \neq R_{m+2}$, then \textbf{return} $y \leftarrow y + b\left(\frac{R_{m+k+1} - x}{R_{m+k+1} - R_{m+2}}\right)$
		\end{enumerate}
	\item elseif $R_{m+1} \le x$ 
		\begin{enumerate}
			\item if $R_{m+2} < R_n$ and $x < R_{m+2}$, then \textbf{return} $y \leftarrow 1$, else \textbf{return} $y \leftarrow 0$
		\end{enumerate}
	\item else
		\begin{enumerate}
			\item if $R_{m+1} \le x$ and $R_{m+2} > R_n$, then \textbf{return} $y \leftarrow 1$, else \textbf{return} $y \leftarrow 0$
		\end{enumerate}				
\end{enumerate}
\end{algo}

\begin{figure*}[h!]
	\centering
	\includegraphics[width=0.65\textwidth]{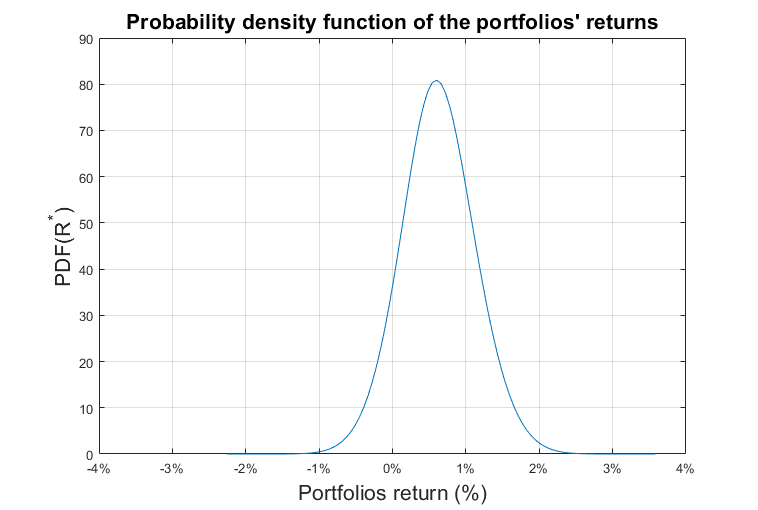}	
	\caption{Illustration of the PDF of the portfolios returns, when the asset returns are given as in Table~\ref{tab:Illustration_Score}.}
	\label{tab:IllustScore2}
\end{figure*}

\noindent As an illustration, let us consider the previous example with 10 assets.
Using Algorithm \ref{algo_CoxDeBoor}, we compute the PDF of the portfolios' returns and report it in Figure~\ref{tab:IllustScore2}.
In this example, we observe that the distribution is uni-modal with most portfolios having a return close to $0.6\%$.

Regarding the computation time, we consider markets of $10$ and $20$ assets whose returns are randomly drawn from a standard normal distribution.
We report in Table~\ref{tab:CompTimeCoxDeBoor} the average computation time and its standard deviation when we compute the PDF of the portfolio returns at the cross-sectional average of the asset returns, i.e. $R^*=\bar{R}$.
The computation is repeated 1000 times.

\begin{table}[h!]\centering
\begin{tabular}{|l|cc|}\hline
Number of Assets 		& 10 & 20 \\
\hline\hline
Mean computation time	& 2.91e-4	& 0.3050 \\
Standard deviation		& 2.65e-5	& 0.0122\\
\hline
\end{tabular}
\caption{Mean computation time in seconds and standard deviation of computing the PDF at a point for markets of $10$ and $20$ assets. The computations were performed using Matlab\copyright \ on a bi-xeon E2620 v3 under Windows\copyright.}
\label{tab:CompTimeCoxDeBoor}
\end{table}

It should be noted that its computation time prevents its use when the number of assets is above around 20. 





\subsection{Closed form expression}

It is straightforward to get the PDF by deriving Equation~(\ref{eq:Lasserre_CDF}).
Using the same notation as in Section \ref{sec_CF_CDF}, this leads to the following closed form formula for the PDF:
\begin{equation} \begin{array}{rl}
f(R^*, \bm{R}) & = (n-1) \sum\limits_{i \in I} \frac{(R^*-S_i)_+^{n-2}}{\prod\limits_{j=1, j \neq i}^n (S_j - S_i)} \, + \\
        & + \sum\limits_{i \in K} \left(\sum\limits_{j=0}^{m_i-1} (-1)^{j+m_i+1} (n-j-1)\binom{n-1}{j} \frac{(R^*- S_i)_+^{n-j-2}}{\prod\limits_{k \in S \setminus \{S_i\}} (S_k -S_i)} \Phi_{m_i-1-j}(\bm{b}_{i}) \right) . \\
\end{array}
\label{eq:Lasserre_PDF}
\end{equation}

Its computation suffers from the same drawback as that of the CDF, providing numerically unstable results for $n\geq 20$ at the usual machine precision.

\subsection{Numerical derivation}

The iterative nature of the de Boor-Cox formula makes the computation of the PDF slow for large number of assets, say $\geq 20$, and its computation using the closed form formula above is numerically unstable for $n\geq 20$.
So, an alternative is to approximate the PDF by deriving numerically the CDF obtained earlier, using Varsi's algorithm, by finite differences.

Let $F$ be the CDF and $x_0$ the point in which we wish to estimate its derivative.
One may employ central differences and the five points' method, thus having
\begin{equation}\label{eq:apr_PDF}
F'(x_0) = \frac{-F(x_0+2h)+8F(x_0+h)-8F(x_0-h)+F(x_0-2h)}{12h} + \frac{h^4}{30} F^{(5)}(c),
\end{equation}
with $c \in [x_0-2h, x_0+2h]$.
The truncation error is then $O(h^4)$.
Even though it is only an estimate of the PDF, this approach enables us to scale up to thousands of assets with computation times being~5 times higher than those reported 
in Table~\ref{tab:CompTime_CDF} for an estimate at a single point with the five points method.
We note that one may approximate the PDF with arbitrarily small error by using higher order differences.





\section[Moments of portfolios' returns distribution]{Moments of the portfolios' returns distribution} \label{sec:Moments}

In this section, we compute the moments of the portfolios returns. 
In Section~\ref{Bary}, we provide affine maps to pass from barycentric to Cartesian coordinates, and vice-versa, and in Section~\ref{Moments_def}, we state the moments' definitions in Cartesian coordinates.
In Section~\ref{Lasserre}, we recall a theorem proposed in \cite{LA2001} which provides an elegant expression for the integral of a symmetric $q$-linear form on a simplex, while in Section \ref{Identities}, we offer identities for some of the nested sums used in the theorem.
In Section \ref{Ind_Terms}, we compute the individual terms that make up the moments up to the fourth one, and in Section~\ref{Moments_ClosedForm}, we provide their closed form solutions.
Finally, in Section \ref{Moments_Algo}, we propose a general algorithm to compute any moment.

\subsection{Barycentric and Cartesian representations}\label{Bary}

\noindent There are affine maps to pass 
\begin{itemize}
	\item from barycentric to Cartesian coordinates: 
		\begin{equation}\label{eq:2}
		\begin{array}{rccl}
		m_{bc}: & \mathbb{R}^{n} & \rightarrow & \mathbb{R}^{n-1},\\
        & x & \mapsto & \breve{x} = T \tilde{x} + v_n,\\
		\end{array}
		\end{equation}
		where $\tilde{x} = (x_i)_{i=1,\dots,n-1}$ , the simplex vertices $v_i$ are expressed in Cartesian coordinates, i.e. $(n-1)$-dimensional column vectors, and
		$T = \left[\begin{array}{c} v_1 -v_n, \  \cdots, \ v_{n-1} -v_n \end{array}\right]$ is an $(n-1)\times (n-1)$ matrix.
	\item from Cartesian to barycentric coordinates:
		\begin{equation}\label{eq:3}
		\begin{array}{rccl}
		m_{cb}: & \mathbb{R}^{n-1} & \rightarrow & \mathbb{R}^{n}, \\
            & \breve{x} & \mapsto & x = \left[\begin{array}{c} I_{n-1}\\ - 1_{n-1}' \end{array}\right] T^{-1} (\breve{x}-v_n) + \left[\begin{array}{c} 0_{n-1}  \\ 1\end{array}\right],
		\end{array}
		\end{equation}
		where $0_{n-1}$ and $-1_{n-1}$ are the $(n-1)$-dimensional column vectors of 0's and $-1$'s, respectively, $I_{n-1}$ is the $(n-1)\times (n-1)$ identity matrix.
\end{itemize}

For individual asset returns $\bm{R} =(R_1, \dots, R_n)$, the return of portfolio $x$ is then given by:
\begin{equation}
R'x = A\breve{x} - A v_n + R_n ,
\end{equation}
where $A = R'\left[\begin{array}{c} I_{n-1}\\ - 1_{n-1}' \end{array}\right] T^{-1}$.


\subsection{Moments}\label{Moments_def}


By definition, the moments of the portfolio returns distribution are given as follows, where $V(\cdot)$ denotes volume:
\begin{equation}\label{eq:4}
M_1 = \frac{1}{V\left(\Delta^{n-1}\right)} \displaystyle\int\limits_{\Delta^{n-1}} A\breve{x} - A v_n + R_n \ d\breve{x},
\end{equation}

$$
M_2 = \frac{1}{V\left(\Delta^{n-1}\right)} \displaystyle\int\limits_{\Delta^{n-1}} (A\breve{x} - A v_n + R_n - M_1)^2 \ d\breve{x},
$$

$$
M_k = \frac{1}{V\left(\Delta^{n-1}\right) (\sqrt{M_2})^k} \displaystyle\int\limits_{\Delta^{n-1}} (A\breve{x} - A v_n + R_n - M_1)^k \, d\breve{x} ,\, k\geq 3 ,
$$
where the term $\frac{1}{V\left(\Delta^{n-1}\right)}$ is normalizing the equations. Indeed, the distance between the vertices $v_i$ is arbitrary, and so is the volume of $\Delta^{n-1}$. 
An alternative is to choose the distance between the vertices $v_i$ such that $V\left(\Delta^{n-1}\right)=1$.
Note that by construction we have $V\left(\Delta^{n-1}\right) = \displaystyle\int\limits_{\Delta^{n-1}} 1 \ d\breve{x}$.

\subsection{Integrating over $\Delta^{n-1}$}
\label{Lasserre}
From \cite{LA2001}, and slightly adapted to our notation, we have the following

\begin{thm}\label{thm_Lass}
Let $v_1, \dots, v_n$ be the vertices of an $(n-1)-$dimensional simplex $\Delta^{n-1}$. Then, for a symmetric q-linear form $H: (\mathbb{R}^{n-1})^q \rightarrow \mathbb{R}$, we have 
\begin{equation}
\displaystyle\int\limits_{\Delta^{n-1}} H(X,\dots,X) \ d\breve{x} = \frac{V\left(\Delta^{n-1}\right)}{\left( \begin{array}{c} n-1+q\\ q \end{array} \right)} \displaystyle\sum\limits_{1\leq i_1 \leq i_2 \leq \dots \leq i_q \leq n} H(v_{i_1}, v_{i_2}, \dots, v_{i_q}),
\end{equation}
\end{thm}

\subsection{Nested sum identities}\label{Identities}

Since Theorem~\ref{thm_Lass} makes use of nested sums, we shall employ the following identities:
\begin{lem}\label{Lemma_id_1}
For $n \in \NN$, it holds 
\begin{equation}
2 \sum_{i=1}^n \sum_{j=i}^n x_i x_j = \left( \sum_{i=1}^n x_i \right)^2 + \sum_{i=1}^n x_i^2.
\end{equation}
\end{lem}

\noindent \textit{Proof}.
Straightforward.

\begin{lem}\label{Lemma_id_2}
For $n \in \NN$, it holds 
$$
(3!) \sum_{i=1}^n \sum_{j=i}^n \sum_{k=j}^n x_i x_j x_k = \left( \sum_{i=1}^n x_i \right)^3 + 3 \left( \sum_{i=1}^n x_i \right) \left( \sum_{i=1}^n x_i^2 \right) + 2 \sum_{i=1}^n x_i^3.
$$
\end{lem}

See proof in Annex \ref{Ax5}.

\begin{lem}\label{Lemma_id_3}
For $n \in \NN$, it holds 
\begin{align*}
  & (4!) \sum_{i=1}^n \sum_{j=i}^n \sum_{k=j}^n \sum_{t=k}^n x_i x_j x_k x_t =\\
= & \left( \sum_{i=1}^n x_i \right)^4 + 6 \left( \sum_{i=1}^n x_i \right)^2 \left( \sum_{i=1}^n x_i^2 \right) + 8 \left( \sum_{i=1}^n x_i \right) \left( \sum_{i=1}^n x_i^3 \right) + 6 \left( \sum_{i=1}^n x_i^4 \right) + 3 \left( \sum_{i=1}^n x_i^2 \right)^2.
\end{align*}
\end{lem}

See proof in Annex \ref{Ax6}.

\subsection{Expression of moments' individual terms}
\label{Ind_Terms}

First, let us note that we have the following identity by construction:

\begin{lem}\label{Lemma_id_Bary0}
For $n \in \NN$, vertices $(v_i)_{i=1}^n$, and matrix $A$ as defined previously, it holds 
\begin{equation}
\sum\limits_{i=1}^n Av_i = A \sum\limits_{i=1}^n v_i = 0.
\end{equation}
\end{lem}

As we shall see in Section~\ref{Moments_ClosedForm}, the moments can be expressed in terms of $\int_{\Delta^{n-1}} (A\breve{x})^p \ d\breve{x}$ with $p \in \NN$.
So, to compute the first four moments, we compute this term for $p \in \{1,2,3,4\}$.

\begin{lem}\label{Lemma_id_X}
For $n \in \NN$, it holds 
\begin{equation}
\int_{\Delta^{n-1}} A\breve{x} \ d\breve{x} = 0 .
\end{equation}
\end{lem}

\noindent \textit{Proof}.
Apply Theorem~\ref{thm_Lass}, with $q=1$ and $H(\breve{x_1}) = A\breve{x_1}$, then it suffices to recall Lemma~\ref{Lemma_id_Bary0}.

\begin{lem}\label{Lemma_id_4}
For $n \in \NN$, it holds 
\begin{equation}
\int_{\Delta^{n-1}} (A\breve{x})^2 \ d\breve{x} = \frac{V\left(\Delta^{n-1}\right)}{n (n+1)} \sum_{i=1}^n (A v_i)^2 .
\end{equation}
\end{lem}

\noindent \textit{Proof}.
Apply Theorem~\ref{thm_Lass}, with $q=2$ and $H(\breve{x_1}, \breve{x_2}) = (A\breve{x_1})(A\breve{x_2})$, by replacing the nested sum in the theorem as in Lemma \ref{Lemma_id_1}, then it suffices to recall Lemma~\ref{Lemma_id_Bary0}.

\begin{lem}\label{Lemma_id_5}
For $n \in \NN$, it holds 
\begin{equation}
\displaystyle\int\limits_{\Delta^{n-1}} (A\breve{x})^3 \ d\breve{x} = \frac{2 V\left(\Delta^{n-1}\right)}{n(n+1)(n+2)} \sum_{i=1}^n (A v_i)^3 .
\end{equation}
\end{lem}

\noindent \textit{Proof}.
Apply Theorem \ref{thm_Lass}, with $q=3$, and $H(\breve{x_1}, \breve{y_2}, \breve{y_3}) = (A\breve{x_1})(A\breve{x_2})(A\breve{x_3})$, by replacing the nested sum in the theorem as in Lemma \ref{Lemma_id_2}, then it suffices to recall Lemma~\ref{Lemma_id_Bary0}.

\begin{lem}\label{Lemma_id_6}
For $n \in \NN$, it holds 
\begin{align*}
\displaystyle\int\limits_{\Delta^{n-1}} (A\breve{x})^4 \ d\breve{x} = & \frac{ V\left(\Delta^{n-1}\right)}{n(n+1)(n+2)(n+3)} \left( 6 \left( \sum_{i=1}^n (A v_i)^4 \right) + 3 \left( \sum_{i=1}^n (A v_i)^2 \right)^2 \right) .
\end{align*}
\end{lem}

\noindent \textit{Proof}.
Apply Theorem \ref{thm_Lass}, with $q=4$, and $H(\breve{x_1}, \breve{y_2}, \breve{y_3}, \breve{y_4}) = (A\breve{x_1})(A\breve{x_2})(A\breve{x_3})(A\breve{x_4})$, replacing the nested sum in the theorem as in Lemma \ref{Lemma_id_3}, then it suffices to recall Lemma~\ref{Lemma_id_Bary0}.

\subsection{Closed form expression of the first four moments}\label{Moments_ClosedForm}

In this section, we derive the closed form expressions for the first four moments, reported in Theorems \ref{thm_M1} to~\ref{thm_M4}, respectively.

First, let us note that by construction we have the identity:
\begin{lem}\label{Lemma_id_Bary1}
For $n \in \NN$, $(R_i)_{i=1}^n$, $(v_i)_{i=1}^n$, and $A$ as defined previously, it holds 
\begin{equation}
R_i = Av_i - Av_n + R_n, \; i \in \left\{ 1, \dots, n\right\}.
\end{equation}
\end{lem}

Next, we show that the first moment of the portfolio returns distribution is equal to the first moment of asset returns as stated in Theorem~\ref{thm_M1}.

\begin{thm}\label{thm_M1}
In a market of $n$ assets, $n \in \NN$, whose returns are $\bm{R} = (R_i)_{i=1}^n$, the first moment of the portfolios' returns is
\begin{equation}
M_1 = \frac{1}{n} \displaystyle\sum\limits_{i=1}^n R_i .
\end{equation}
\end{thm}

\noindent \textit{Proof}.
Develop $M_1$ in Equation~\ref{eq:4} and simplify it using Lemma~\ref{Lemma_id_X}, then apply Lemma~\ref{Lemma_id_Bary1}.

From Theorem \ref{thm_M1} and Lemma~\ref{Lemma_id_Bary1}, we get the identity: 

\begin{lem}\label{Lemma_id_Bary2}
For $n \in \NN$, $(R_i)_{i=1}^n$, $(v_i)_{i=1}^n$, $A$ and $M_1$ as defined previously, it holds 
\begin{equation}
M_1 = R_n - Av_n.
\end{equation}
\end{lem}

It follows that, by employing Lemma \ref{Lemma_id_Bary2}, $M_2$ and $M_k$ simplify to

\begin{equation}\label{eq:5}
M_2 = \frac{1}{V\left(\Delta^{n-1}\right)} \displaystyle\int\limits_{\Delta^{n-1}} (A\breve{x})^2 \ d\breve{x},
\end{equation}

\begin{equation}\label{eq:6}
M_k = \frac{1}{V\left(\Delta^{n-1}\right) (\sqrt{M_2})^k} \displaystyle\int\limits_{\Delta^{n-1}} (A\breve{x})^k d\breve{x} ,\, k\geq 3 .
\end{equation}

From Lemma~\ref{Lemma_id_Bary1} and Lemma~\ref{Lemma_id_Bary2}, we have
\begin{lem}\label{Lemma_id_Bary3}
For $n \in \NN$, $(R_i)_{i=1}^n$, $(v_i)_{i=1}^n$, $A$ and $M_1$ as defined previously, it holds 
\begin{equation}
Av_i = R_i - M_1.
\end{equation}
\end{lem}
\noindent
which is used to compute the second, third and fourth moments as stated in Theorems \ref{thm_M2} to~\ref{thm_M4}

\begin{thm}\label{thm_M2}
In a market of $n$ assets, $n \in \NN$, whose returns are $\bm{R} = (R_i)_{i=1}^n$, the second moment of the portfolios' returns is
\begin{equation}
M_2	= \frac{1}{n (n+1)} \sum_{i=1}^n (R_i - M_1)^2 = \frac{1}{n+1} Var(\bm{R}),
\end{equation}
where $Var(\bm{R})=\frac{1}{n} \sum\limits_{i=1}^n (R_i - M_1)^2$ is the (biased) sample variance of the asset returns.
\end{thm}

\noindent \textit{Proof}.
Replace $\displaystyle\int\limits_{\Delta^{n-1}} (A\breve{x})^2 \ d\breve{x}$ \, in Equation~(\ref{eq:5}) by the expression from Lemma~\ref{Lemma_id_4}, then apply Lemma \ref{Lemma_id_Bary3}.
 
\begin{thm}\label{thm_M3}
In a market of $n$ assets, $n \in \NN$, whose returns are $\bm{R} = (R_i)_{i=1}^n$, the third moment of the portfolios' returns is
\begin{equation}
M_3 = \frac{1}{M_2^{3/2}} \frac{2}{n (n+1) (n+2)} \sum_{i=1}^n (R_i - M_1)^3 = \frac{2 \sqrt{n+1}}{n+2} Skew(\bm{R}),
\end{equation}
where $Skew(\bm{R}) = \frac{1}{Var(\bm{R})^{3/2}} \frac{1}{n} \sum\limits_{i=1}^n (R_i - M_1)^3$ is the Fisher-Pearson coefficient of skewness  of the asset returns.
\end{thm}
 
\noindent \textit{Proof}.
By replacing $\displaystyle\int\limits_{\Delta^{n-1}} (A\breve{x})^3 \ d\breve{x}$ in Equation~(\ref{eq:6}) with the expression from Lemma \ref{Lemma_id_5}, and applying Lemma \ref{Lemma_id_Bary3}, one obtains the result.

\begin{thm}\label{thm_M4}
In a market of $n$ assets, $n \in \NN$, whose returns are $\bm{R} = (R_i)_{i=1}^n$, the fourth moment of the portfolios' returns is
\begin{equation}
\begin{array}{rl}
M_4 & = \frac{1}{M_2^2} \frac{1}{n (n+1) (n+2) (n+3)} \left( 6 \sum\limits_{i=1}^n \left(R_i-M_1\right)^4 + 3 \left( \sum\limits_{i=1}^n \left(R_i-M_1\right)^2 \right)^2 \right)\\
    & = \frac{3 (n+1)}{(n+2) (n+3)} \left( 2 Kurt(\bm{R}) + n \right)\\
\end{array}
\end{equation}
where $Kurt(\bm{R}) = \frac{1}{Var(\bm{R})^{2}} \frac{1}{n} \sum\limits_{i=1}^n (R_i - M_1)^4 $ is the Kurtosis of the asset returns.
\end{thm}

\noindent \textit{Proof}.
By replacing $\displaystyle\int\limits_{\Delta^{n-1}} (A\breve{x})^4 \ d\breve{x}$ in Equation~(\ref{eq:6}) with the expression from Lemma \ref{Lemma_id_6}, and applying Lemma \ref{Lemma_id_Bary3}, the claim is established.

To conclude this section, we find a direct mapping between the first four moments of the distribution of portfolio returns (PRD) and those of the cross-sectional asset returns distribution (ARD). 
The first moments of each of these distributions are equal.
Their second and third moments are proportional to each other by a factor depending on the number of assets. Notice that both factors in Equations~(\ref{thm_M2}), (\ref{thm_M3}) are $\leq 1$.
Thus, the second moment of the PRD is smaller 
than the respective moment of the ARD;
the same inequality holds for the absolute values of the respective third moments. 
The fourth moment of PRD is an affine transformation of ARD's fourth moment.
One may observe that the larger the number of assets, the more mesokurtic the PRD gets whatever the fourth moment of ARD is.
Furthermore, focusing on PRD, we highlight the {\em kurtosis} of the asset returns $\bm{R}$,
namely $Kurt(\bm{R})>\frac{14}{3}$ $M_4 >3$, i.e.\ PRD is leptokurtotic.

\subsection{General algorithm to compute the moments} \label{Moments_Algo}

The problem here is to compute the $k^{th}$ moment of the portfolios' returns, i.e., by restating Equation~(\ref{eq:6}), one computes
\[
M_k = \frac{1}{Vol\left(\Delta^{n-1}\right) (\sqrt{M_2})^k} \displaystyle\int\limits_{\Delta^{n-1}} (A\breve{x})^k d\breve{x}\ ,\, k\geq 3 .
\]



Let us set $H(X_1, \dots, X_k)=\prod\limits_{i=1}^k X_i$. It is a symmetric $k$-linear form, and we have 
\[
H(\underbrace{A\breve{x}, \dots, A\breve{x}}_{k \text{ times}}) = (A\breve{x})^k .
\]
Thus, following from Theorem~\ref{thm_Lass}, we obtain
\[
\displaystyle\int\limits_{\Delta^{n-1}} H(\underbrace{A\breve{x}, \dots, A\breve{x}}_{k \text{ times}}) \ d\breve{x} = \frac{V\left(\Delta^{n-1}\right)}{\left( \begin{array}{c} n-1+k\\ k \end{array} \right)} \displaystyle\sum\limits_{1\leq m_1 \leq m_2 \leq \dots \leq m_k \leq n} H(Av_{m_1}, Av_{m_2}, \dots, Av_{m_k}).
\]

Let $q_i$ be the number of occurrences of value $i$, $1\leq i\leq n$. Then,
\[
\displaystyle\int\limits_{\Delta^{n-1}} \left(A\breve{x}\right)^k \ d\breve{x} = \frac{V\left(\Delta^{n-1}\right)}{\left( \begin{array}{c} n-1+k\\ k \end{array} \right)} \displaystyle\sum\limits_{\sum\limits_{i=1}^n q_i = k} \prod\limits_{i=1}^n \left(Av_{i}\right)^{q_i} .
\]
Now, we change the notations. Let $\lambda=\left(\lambda_1, \dots,\lambda_k\right)$ be a partition of $k$, and $\Lambda_k$ the set of partitions of $k$.
We denote by $l=(l_i)_{i=1}^d$ the $d$ unique non-zero values in $\lambda$, $d\leq k$, and by $(p_i)_{i=1}^d$ the multiplicities of $(l_i)_{i=1}^d$. For instance, $\lambda = (2,1,1,0)$ is a partition of $k=4$, with $d=2$, $l=(1,2)$ and the associated multiplicities $p=(2,1)$.\\

From~\cite[Eq. (2.14')]{MacDonald1995}, we have
\[
\displaystyle\sum\limits_{\sum\limits_{i=1}^n q_i = k} \prod\limits_{i=1}^n \left(Av_{i}\right)^{q_i} = \displaystyle\sum\limits_{\lambda \in \Lambda_k} \frac{\prod\limits_{i=1}^d \left( \sum\limits_{j=1}^n (Av_j)^{l_i}\right)^{p_i}}{\prod\limits_{i=1}^d p_i! l_i^{p_i}}.
\]

Set $\Lambda_k$ can be obtained by Algorithm ZS1 in~\cite{ZS1994}, and is still tractable for large moments, as shown in Table~\ref{tab:ZS1}.

\begin{table}[htbp] \centering
\begin{tabular}{|l||cccccc|}\hline
$k$	& 1 & 5 & 10 &  20 &   30 & 40\\
\hline\hline
$|\Lambda_k|$ & 1 & 7 & 42 & 627 & 5604 & 37338\\\hline
\end{tabular}
\caption{Number of partitions $|\Lambda_k|$ for different values of $k$.\label{tab:ZS1}}
\end{table}


The computation is formally presented in Algorithm~\ref{algo_Moments} below.
As an illustration of the computation times, we compute the $k^{\rm th}$ order moments, $k=5,10,15,20$, for markets of $100$, $1000$ and $10,000$ assets whose returns are randomly drawn.
The computation is repeated 1000 times. We report in Table \ref{tab:CompTimeMoments} the average computation time in seconds and its standard deviation.

\begin{algo}\label{algo_Moments}
Let $\bm{R}$ be the asset returns, $N$ the number of assets, and $k$ the moment order.
\begin{enumerate}
	\item Compute $M_2$ by Theorem \ref{thm_M2}
	\item Compute $Av$ as in Lemma \ref{Lemma_id_Bary3}
	\item Compute $\Lambda$ using Algorithm ZS1 in \cite{ZS1994}
	\item Set $S=0$
	\item For each $\lambda \in \Lambda$:
		\begin{enumerate}
			\item decompose $\lambda$ in its $d$ non-zero elements $(l_i)_{i=1}^d$ with multiplicities $(p_i)_{i=1}^d$
			\item $a = \prod\limits_{i=1}^d p_i!\, l_i^{p_i}$  
			\item $b = \prod\limits_{i=1}^d \left( \sum\limits_{j=1}^n (Av_j)^{l_i}\right)^{p_i}$
			\item $S = S + a/b$
		\end{enumerate}
	\item Set $M_k = S\, / \, \left( \sqrt{M_2}^k \cdot \left( \begin{array}{c} n-1+k\\ k \end{array} \right)\right)$	
\end{enumerate}
\end{algo}

\begin{table}[h!]\centering
\begin{tabular}{|l||cccc|}\hline
Moment order:	& 5	&	10	& 15	& 20		\\
Nb of Assets	&   	&		& 	&		\\
\hline\hline
100    		&  0.0006   &  0.0024  &  0.0100  &  0.0398		\\
		& (0.0021)  & (0.0003) & (0.0005) & (0.0005)	\\
\hline	
1000		&  0.0008   &  0.0034  &  0.0117  &  0.0420		\\
		& (0.0000)  & (0.0003) & (0.0002) & (0.0007)	\\
\hline	
10,000		& 0.0020    &  0.0053  &  0.0145  &  0.0506		\\
		& (0.0001)  & (0.0000) & (0.0011) & (0.0028)	\\\hline			
\end{tabular}
\caption{Mean runtime in seconds, and standard deviation in parenthesis, of computing the $k$-th order moment in markets of up to $10,000$ assets. 
Asset returns are drawn randomly before each of the 1000 computations.
Experiments performed with Matlab\copyright \ on a bi-xeon E2620 v3 under Windows\copyright.
\label{tab:CompTimeMoments}}
\end{table}

\section{Applications of the score in asset allocation}\label{sec:dist_score}

In this section, we illustrate the usefulness of the score in asset allocation.
The main method to perform an asset allocation is by mean-variance optimization.
Introduced by \cite{Markowitz56}, mean-variance efficiency is an optimal portfolio construction which quantifies the value of diversification.
It gives more (less) weight to the assets with large (small) estimated returns, negative (positive) correlations and small (large) variances.
Despite its theoretical foundation, asset managers did not adopt this method very quickly, see \cite{M1989}.
Indeed, with sample moments, the method leads to optimal portfolios which are both very concentrated and presenting a high turn-over.
As argued in \cite{M1989}, this is due to estimation errors in the moments and to the method which acts as an "estimation-error maximizer".
Several ways have been proposed to overcome this issue.
The main ones are to get more robust estimates (see \cite{LW04}, \cite{HLPL2006} among others), to constrain the optimal portfolios (see e.g. \cite{DMGNU2009}) and to propose alternative objective functions such as risk-parity portfolios (see \cite{MRT2010}).

Here, by estimating the distribution of a portfolio’ score, we quantify the risk for this portfolio to be among the worst (and best) performers. 
It introduces a new concept of risk, based on portfolio ranking, which shall matter to asset managers caring about the relative performance of their allocation.\footnote{For instance, a newly established asset manager wants to attract funds and he is likely to worry about performing poorly compared to his peers.}
Consequently, we propose new performance measures based on a portfolio score and we use these measures in a portfolio optimization problem.
We compare our optimal portfolios with those obtained by mean-variance (MV) optimization using the shrinkage estimator of \cite{LW04} and by risk parity (ERC), as in \cite{MRT2010}, also using the shrinkage estimator.
The shrinkage estimator is probably one of the most used to overcome the estimation errors in the covariance matrix. 
We compare these optimal portfolios in a pseudo-real time asset allocation exercise to assess their out-of-sample performance.
Overall, we find that the score-based optimal portfolios perform better than the mean-variance portfolio and that they hold the comparison with the ERC portfolio.
We find that the score-based optimal portfolios are less concentrated than the mean-variance (MV) optimal portfolio, much less risky in terms of ranking and they present less turn-over.
With respect to the ERC portfolio, they provide similar results but with a range of different concentration and associated turn-over.
These exercises suggest that the score can help in overcoming the issues of concentration and turn-over, and present an interest in asset allocation. 

The section is structured as follow.
First, through two simple examples, we show that the score brings a new piece of information about the portfolios.
Next, we use the score distribution to sketch new performance measures. 
Used in a portfolio optimization problem, these measures lead to new optimal portfolios that are less concentrated than the mean-variance optimal portfolio and much less risky in terms of ranking.
Finally, the score-based optimal portfolios are used in an asset allocation exercise and compared with the MV and ERC portfolios.

\subsection{Portfolio's distribution of score}\label{subsec:dist_score_examples}

Once a portfolio is chosen and assuming a distribution for the asset returns, one can estimate the distribution of the scores of this portfolio.
This distribution allows us to understand the risk for this portfolio to perform worse, or better, than the other portfolios.
This estimation is obtained as follow.
First, we draw randomly $10^7$ vectors of asset returns. 
Then we compute the corresponding scores using Varsi's algorithm (Algorithm \ref{algo_Varsi}).
Finally, we estimate the distributions of score by a normal kernel function bounded in $[0,1]$.
Equipped with this estimation, we present two results.
We show that different portfolios with identical normal return distributions can exhibit very different score distributions.
We also show that the uneven introduction of skewness in normal and exchangeable asset returns' distributions leads to an asymmetry in the portfolios' distributions of scores and to differences in their means.
Thus the score offers a new way to distinguish portfolios. 


In Markowitz' framework, an asset manager cannot choose between portfolios when their returns have the same mean and variance.
In this context, we show that the distribution of the portfolios' score can be very different one another and that a manager caring about his portfolio's ranking would prefer one of them because of its score's variance.
Let consider asset returns following a normal distribution with zero mean and covariance matrix $\Sigma$.\footnote{To use a realistic covariance matrix we estimate $\Sigma$ on the monthly returns of the 10 industry indices of French's dataset ranging from Jan.\ 2000 to Dec.\ 2009 (120 observations). French's dataset is at \url{https://mba.tuck.dartmouth.edu/pages/faculty/ken.french/data_library.html}. The estimation is performed using the shrinkage estimator in~\cite{LW04}. The covariance matrix $\Sigma$ is provided in Appendix~\ref{subSec:EstCovMatrix}.}
And let choose four portfolios having the same volatility, $4.47\%$,\footnote{i.e. a variance of $0.002$.} but quite different one another.
This difference is expressed both in terms of distance between one another and in terms of concentration.\footnote{The distance used is to the $L_1$ distance, also called turn-over distance. It corresponds to the minimum change in weights to pass from a portfolio to another. It is expressed in percentage ranging from 0\% to 200\%. And, we measure the concentration of a portfolio as its $L_1$ distance to the equally weighted portfolio. The larger the distance the more concentrated the portfolio. These differences are reported in Appendix~\ref{subsec:dif_ptfs}.}
These portfolios are reported in Table~\ref{Tab:4Ptfs}. 


\begin{table}[t]
\centering
\footnotesize
\begin{tabular}{l|rrrr}
Asset &   Ptf 1 &   Ptf 2 &   Ptf 3 &   Ptf 4 \\
\hline
   1  &  1.17\% &  3.53\% &  5.45\% & 17.95\% \\
   2  &  0.14\% & 10.99\% &  4.06\% & 10.28\% \\
   3  & 18.83\% & 17.18\% &  9.93\% &  3.96\% \\
   4  & 31.48\% & 20.03\% & 11.77\% & 13.05\% \\
   5  &  3.88\% &  1.12\% &  9.78\% & 11.76\% \\
   6  &  1.02\% &  1.86\% & 11.33\% &  9.03\% \\ 
   7  &  1.75\% &  3.80\% &  3.25\% &  8.73\% \\
   8  &  0.44\% &  9.16\% &  2.67\% &  4.73\% \\
   9  & 39.29\% & 14.27\% & 20.51\% &  6.29\% \\
  10  &  1.98\% & 18.06\% & 21.27\% & 14.23\% \\ 
\end{tabular}
\caption{Four portfolios with the same mean and variance.}
\label{Tab:4Ptfs}
\end{table}

\begin{figure}[t]
\centering
\includegraphics[width=0.65\textwidth]{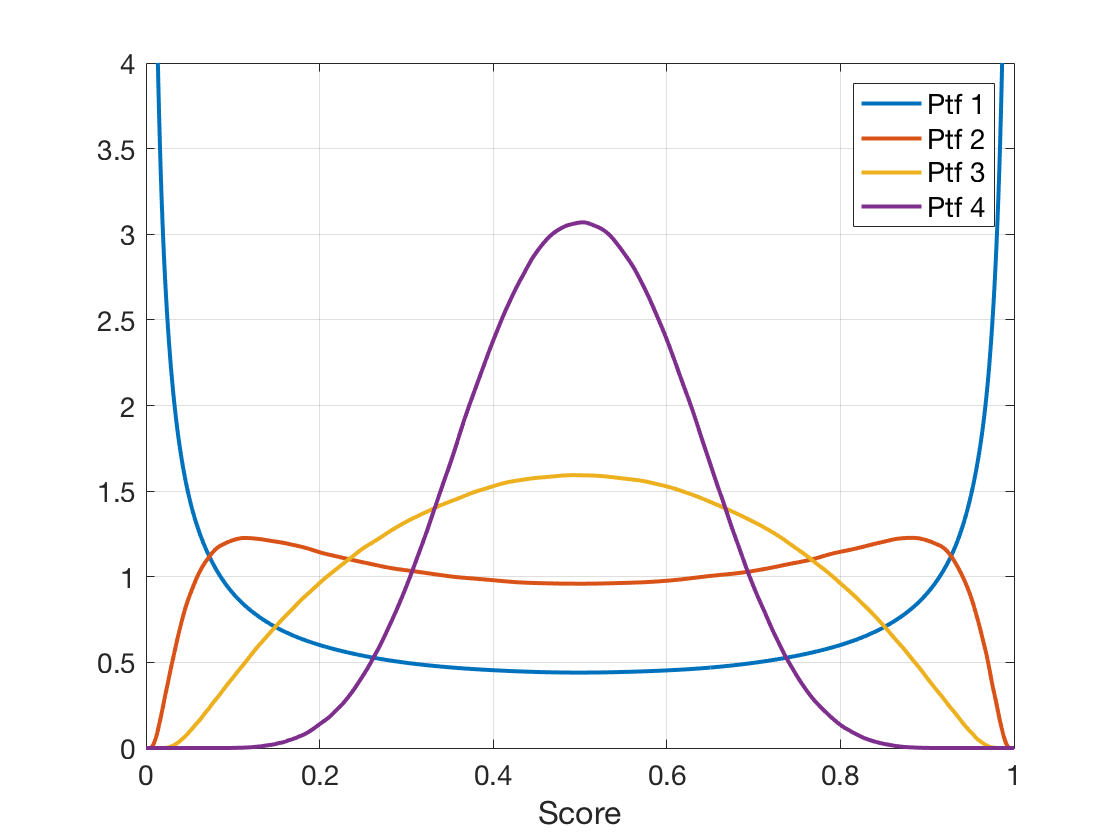}
\caption{Probability density function of the 4 portfolios' scores.}
\label{Fig:PDF4Ptfs_Returns}
\end{figure}

Figure \ref{Fig:PDF4Ptfs_Returns} reports the estimates of the four score distributions.
We observe that they are all symmetric and centered around 0.5, which is expected since the asset returns follow a symmetric distribution centered around the same mean.
However, their dispersion varies a lot and gradually from portfolio 1 whose scores are very dispersed to portfolio 4 whose scores are rather concentrated.
The portfolios having the most dispersed scores are the riskiest ones in terms of ranking.  
An asset manager willing to take risk in terms of portfolio ranking would prefer portfolio 1 whereas a risk-averse (in terms of ranking) manager would prefer portfolio 4. 
In future work, we intend to disentangle the roles of the covariance matrix and of the portfolio concentration in the shape of the score distribution.

The second result is that the uneven introduction of skewness in symmetric and exchangeable asset returns' distribution leads to an asymmetry in the portfolios' distributions of scores and to differences in their means.
Let keep the previous portfolios and assume that the asset returns follow a skew-t distribution.
We suppose that the distribution has mean $\mathbf{0}$, variance-covariance matrix  $0.0035 \ \mathbf{I}$.\footnote{We denote by $\mathbf{0}$ the $10 \times 1$ vectors of 0's and by $\mathbf{I}$ the identity matrix}, and we consider three cases: no skewness, the same skewness and different skewnesses across assets.
The distributions' parameters are reported in Appendix~\ref{subSec:SkewTParameters}.



\begin{figure}[h!]
\centering
\includegraphics[width=1.0\textwidth]{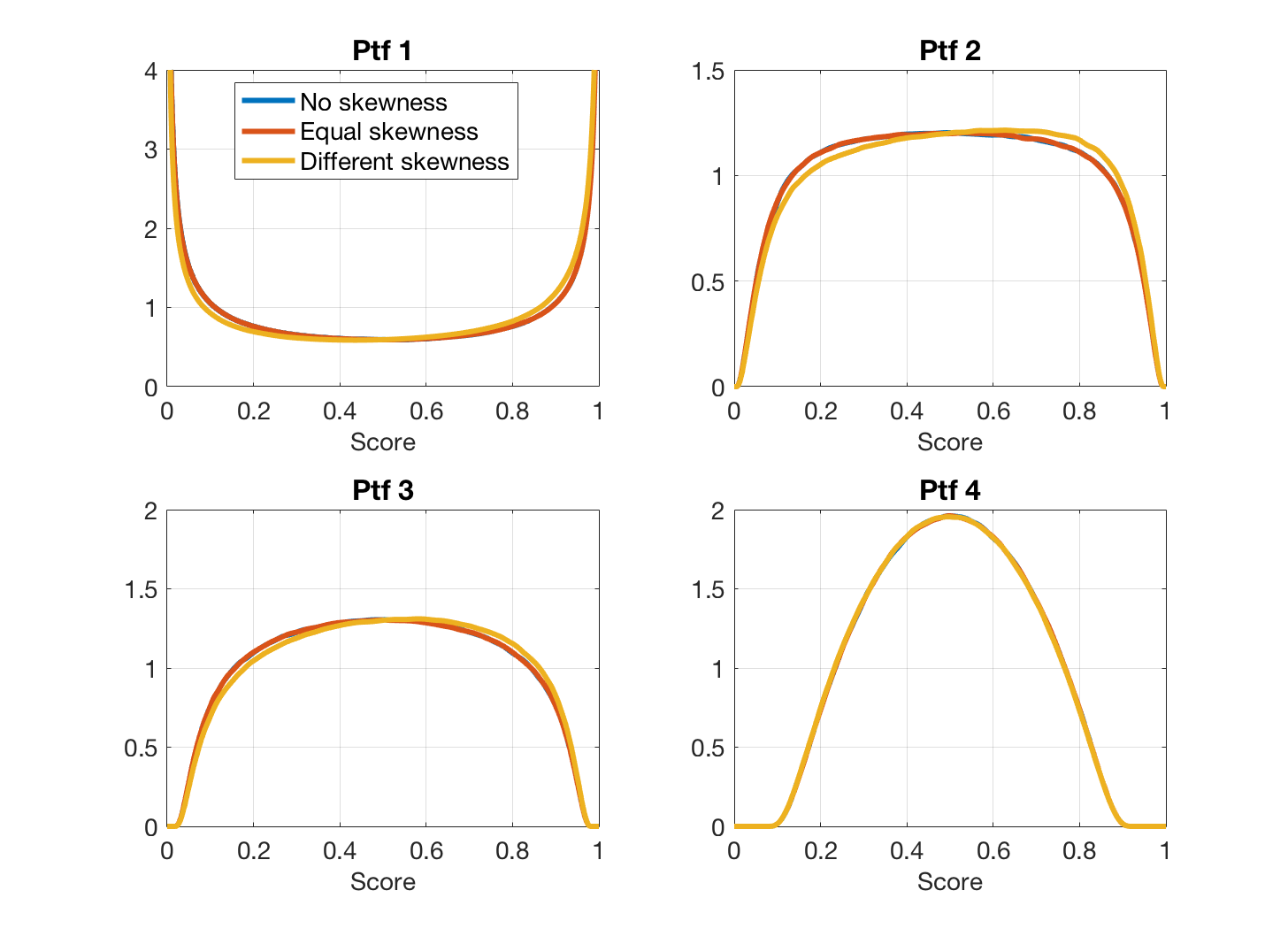}
\caption{PDF of the 4 portfolios' scores in the three cases: no skewness, the same skewness and different skewnesses across assets.}
\label{Fig:PDF4Ptfs_Skewness}
\end{figure}

The distributions of the portfolios' score are presented in Figure \ref{Fig:PDF4Ptfs_Skewness}.
For a better readability we also present their statistics in Table~\ref{Tab:Stats_Ptfs_Scores}.
In the cases of 'no skewness' and 'same skewness', the distributions of the asset returns are exchangeable resulting in distributions of scores which are symmetric and centered around 0.5.
The even introduction of skewness in asset returns does not appear to modify the score distributions.  
When different skewnesses are introduced, we observe an asymmetry in the score distributions.  
In that case, a paired-sample t-test confirms that the average scores are significantly different from $0.5$. 
This skewness favors portfolio 1 whose average score is $53.40\%$.
Here, a manager caring about the ranking of his portfolio might prefer one of them but this time also because of the mean of its score.

\begin{table}[h!]
\footnotesize
\centering
\begin{tabular}{llcccc}
          & Portfolio      &     1   &    2    &    3    &    4     \\
\hline			
No        & Mean           & 50.00\% & 50.01\% & 50.00\% & 50.00\%  \\
skewness  & t-stat (p-val) & 0.34 (0.73) & 1.48 (0.14) & -0.61 (0.54) & 0.47 (0.64) \\
          & Std dev        & 36.13\% & 24.98\% & 23.64\% & 17.16\%  \\
          & Skewness       &  0.00   &  0.00   &  0.00   &  0.00    \\
\hline 			
Same      & Mean      & 50.00\% & 49.99\% & 49.99\% & 50.00\%  \\
skewness  & t-stat (p-val) & 0.11 (0.91) & -1.49 (0.14) & -1.41 (0.16) & -0.73 (0.46)  \\
          & Std dev   & 36.14\% & 24.98\% & 23.64\% & 17.16\%  \\
          & Skewness  &  0.00   &  0.00   &  0.00   &  0.00    \\ 
\hline			
Different & Mean      & 53.40\% & 51.16\% & 51.02\% & 49.91\%  \\ 
skewness  & t-stat (p-val) & 299.07 (0.00) & 147.53 (0.00) & 136.13 (0.00) &-17.05 (0.00) \\
          & Std dev   & 35.97\% & 24.95\% & 23.61\% & 17.17\%  \\
          & Skewness  & -0.15   & -0.05   & -0.04   & 0.00     \\
\hline		
\multicolumn{2}{c}{Exp. Return} & 0\% & 0\% & 0\% & 0\% \\
\multicolumn{2}{c}{Volatility}  & 3.19\% & 2.26\% & 2.21\% & 2.03\% \\
\end{tabular}
\caption{Statistics on the scores and returns of the four portfolios for the different distribution of asset returns. The t-statistics and p-values are of the paired-sample t-test that the mean score is 50\%.}
\label{Tab:Stats_Ptfs_Scores}
\end{table}


In conclusion, those two simple examples show that the score offers a new way to distinguish portfolios which can be interesting to asset managers. 
Also, despite the different asset return distributions in the two exercises, the shape of the score distributions appears rather similar.
In the following, we investigate performance measures based on the score to balance the risk for a portfolio to be among the worst performers and its opportunity to be among the best ones.

\subsection{The performance measures}\label{subsec:performance_measures}

In this section, we propose several performance measures based on the score distribution and we analyse the optimal portfolios associated to these measures.
With respect to the MV portfolio, we find that the score-based optimal portfolios are less concentrated, much less extreme in terms of score and more frequently among the $50\%$ best performers.

First, let introduce some notations.
For a given portfolio, we denote $x$ the random variable of its score, $f$ the probability density function of this score and $S^*$ a given score used as target/benchmark.
The portfolio's expected score is denoted $\mu$ and its variance $\sigma^2$.
We also denote $\mu_{S^*}^+ = \int_{S^*}^1 x f(x) dx$ and $\mu_{S^*}^- = \int_0^{S^*} x f(x) dx$, the mean of the scores which are below and above $S^*$ respectively.

Second, we propose the following performance measures:
\begin{itemize}
	\item Performance measure A: \begin{equation}\label{eq:Perf_A} Perf_A = \frac{ \int_0^1 (x-S^*) f(x) dx }{\sqrt{\int_0^1 (x-S^*)^2 f(x) dx}} = \frac{\mu - S^*}{\sigma} \end{equation} 
	\item Performance measure B: \begin{equation}\label{eq:Perf_B} Perf_B = \frac{ \int_0^1 x f(x) dx }{\sqrt{\int_0^1 x^2 f(x) dx}} = \frac{\mu}{\sigma}\end{equation} 
	\item Performance measure C: \begin{equation}\label{eq:Perf_C} Perf_C = \frac{ \left( \int_{S^*}^1 f(x) dx \right) \left( \int_{S^*}^1 x f(x) dx - S^*\right) }{\left( \int_0^{S^*} f(x) dx \right) \left( S^* - \int_0^{S^*} x f(x) dx\right)} = \frac{ \int_{S^*}^1 f(x) dx }{\int_0^{S^*} f(x) dx } \frac{  \mu_S^+ - S^* }{  S^* - \mu_S^-} \end{equation}
	\item Performance measure D: \begin{equation}\label{eq:Perf_D} Perf_D = \int_0^1 \sqrt{x} f(x) dx \end{equation}
\end{itemize}

$Perf_A$ is a measure analog of the Sharpe ratio, see \cite{Sharpe94}, with the score $S^*$ being the counterpart of the risk-free rate.
This measure increases with $\mu$ increasing and $\sigma$ decreasing. 
However, as with the Sharpe ratio, positive deviation from the mean is seen as a risk.
Also it cannot be used to compare portfolios if their expected scores are equal to $S^*$, whatever the variances of the portfolios' scores.
$Perf_B$ is similar to $Perf_A$ with $S^*=0$. It is harsher than $Perf_A$ with risky portfolios and it can be used to compare portfolios even if their expected scores are identical.
$Perf_C$ puts emphasis on the asymmetry of $f$ above and below $S^*$. 
Finally $Perf_D$ is a simple measure which penalizes the low scores more than it rewards the high scores.

Third, we compute the optimal portfolios associated with these performance measures in two cases of asset return distribution.
In the first exercise we assume that the asset returns follow a multivariate Gaussian distribution with empirical parameters.
The mean is as presented in Table \ref{Tab:ExpectedAR} and the covariance matrix is $\Sigma$.
This is a classical mean-variance portfolio optimization framework and the asymmetry in the score shall come from the different means.
In the second exercise, we assume that the asset returns follow a multivariate Generalized Hyperbolic distribution with mean $\mathbf{0}$, covariance matrix $\Sigma$ and the rest of the parameters fitted to the data, see Appendix \ref{subSec:GHypParameters}.
We choose to set the mean of the asset returns to $\mathbf{0}$ because it is very poorly estimated in practice.
The asymmetry in the score shall come from the asset returns' skewness, as seen in Section \ref{subsec:dist_score_examples}. 

\begin{table}[h!]
\centering
\begin{tabular}{ccccc}
 Asset 1    &   Asset 2    &   Asset 3    &    Asset 4    &    Asset 5   \\
  0.67\%  &  0.17\%    &   0.59\%   &  1.05\%     &  -0.15\%   \\
\hline
Asset 6     &   Asset 7    &   Asset 8    &   Asset 9     &   Asset 10   \\
-0.50\%    &  0.29\%   &  0.33\%   &  0.82\%    &  0.14\%   \\
\end{tabular}
\caption{Empirical mean of the asset returns.
\label{Tab:ExpectedAR}}
\end{table}

In each exercise, we compare the optimal portfolios with the MV optimal portfolio.
We maximize the performance measure by using Sequential Quadratic Programming \cite{sequential2006}.
During optimization, we evaluate the performance measure for different portfolios by estimating the portfolio's PDF of scores as in Section \ref{subsec:dist_score_examples}.
Finally, we constrain the optimal portfolios to have a variance of $0.002$.
The optimal portfolios are in Table \ref{Tab:OptPtfs}.

\begin{table}[h!]
\centering
\footnotesize
\begin{tabular}{l|rrrrr|rrrrr}
          & \multicolumn{5}{c|}{Gaussian} & \multicolumn{5}{c}{ Generalized Hyperbolic} \\ 
Asset     &   MV    &    A    &   B     &   C     &    D    &   MV    &    A    &   B     &   C     &    D   \\
\hline
1         & 26.3\%  & 14.3\%  &  9.1\%  & 13.7\%  & 14.2\%  &  0.0\%  &  16.0\% &   9.2\% &  15.4\% & 16.2\%  \\
2         &  0.0\%  & 11.7\%  & 10.4\%  & 11.4\%  & 11.7\%  & 28.8\%  &  20.7\% &  10.5\% &  19.6\% & 21.3\%  \\
3         &  0.0\%  & 17.5\%  & 10.7\%  & 17.0\%  & 17.5\%  &  0.0\%  &   0.0\% &  10.6\% &   0.0\% &  0.0\%  \\
4         & 53.6\%  & 16.9\%  & 10.0\%  & 16.0\%  & 17.0\%  &  0.0\%  &  15.5\% &  10.0\% &  15.1\% & 15.8\%  \\
5         &  0.0\%  & 10.5\%  & 10.1\%  & 10.9\%  & 10.4\%  &  0.0\%  &   9.0\% &  10.1\% &   9.8\% &  8.5\%  \\
6         &  0.0\%  &  0.0\%  & 10.2\%  &  0.0\%  &  0.0\%  &  0.0\%  &   0.0\% &  10.1\% &   0.0\% &  0.0\%  \\
7         &  0.0\%  &  9.2\%  &  9.7\%  &  9.8\%  &  9.1\%  &  0.0\%  &   1.4\% &   9.6\% &   3.3\% &  0.5\%  \\
8         &  0.0\%  &  4.2\%  &  9.4\%  &  4.9\%  &  4.2\%  & 71.2\%  &  22.1\% &   9.4\% &  20.5\% & 23.0\%  \\
9         & 20.0\%  & 11.3\%  &  9.8\%  & 10.8\%  & 11.5\%  &  0.0\%  &   3.6\% &   9.7\% &   4.1\% &  3.2\%  \\
10        &  0.0\%  &  4.4\%  & 10.7\%  &  5.5\%  &  4.4\%  &  0.0\%  &  11.8\% &  10.7\% &  12.2\% & 11.4\%  \\
\hline
Exp.      &         &         &         &         &         &         &         &         &         &         \\
Ret.      & 0.90\%  & 0.52\%  & 0.34\%  & 0.50\%  & 0.52\%  &   0\%   &   0\%   &    0\%  &    0\%  &   0\%   \\
\end{tabular}
\caption{Mean-variance and score-based optimal portfolios when the asset returns follow the multivariate Gaussian and Generalized Hyperbolic distributions.}
\label{Tab:OptPtfs}
\end{table}

Let consider the case of Gaussian asset returns. 
For readability we report the estimated PDF of the scores in Appendix \ref{subSec:ScoreDist_Gaussian}, some associated probabilities in Appendix \ref{subSec:ProbaScores_Gaussian}, and the distances among the optimal portfolios in Appendix \ref{subSec:PtfsDistances_Gaussian}.
As expected the score distributions are asymmetric.
Focusing on the MV portfolio, we observe that is very concentrated, investing in only 3 assets. 
By construction, it has the largest expected return among the optimal portfolios. 
The density of its score is U-shaped meaning that the portfolio is very likely to be among the very best or the very worst performers.
Indeed, it has a probability of $26.6\%$ to be among the $10\%$ worst performers, a probability of $40.4\%$ to be among the $10\%$ best performers, and a probability of only $21.9\%$ to perform a score between $20\%$ and $80\%$. Overall, it has a probability of $57.6\%$ to be among the top $50\%$ performers.
We also observe that the portfolios based on $Perf_A$, $Perf_C$ and $Perf_D$ are very similar one another.
They are less concentrated than the MV portfolio with less than $50\%$ of their weights in the 3 most invested assets.
The density of their scores has an inverted U-shape, i.e. the portfolios are rarely among the very worst and very best performers.
They also have a probability of $60.2\%$ to be among the top $50\%$ performers.
Finally the portfolio based on $Perf_B$ is very close to the equally weighted portfolio.
Its scores are very steady around $0.5$ and it has the lowest expected return.
Overall, the low concentration and the lower dispersion in terms of score offered by the score-based optimal portfolios come at the cost of half the MV portfolio's expected return. \footnote{This cost can also be observed when plotting the optimal portfolios against the efficient frontier in Appendix \ref{subSec:EfficientFrontier_Gaussian}.}

In the case of Generalized Hyperbolic asset returns, the estimated PDF of the portfolio scores are presented in Appendix \ref{subSec:ScoreDist_GHyp}, some statistics in Appendix \ref{subSec:ProbaScores_GHyp}, and the distances among the optimal portfolios in Appendix \ref{subSec:PtfsDistances_GHyp}.
The expected return of all portfolios is zero by construction.
The conclusions are qualitatively the same as in the previous exercise but with more concentrated portfolios and more extreme score distributions.
The shape of their score distributions is similar to the previous ones.

To conclude, in both cases the score-based portfolios are less concentrated than the MV portfolio and much less risky in terms of score.
The optimal portfolio's score distributions are asymmetric and their shape are quite similar between the two exercises.
Moreover, as exemplified in the second exercise, when using the score, there is no need to estimate the mean of the asset returns to select a portfolio some skewness in the asset returns is enough to cause asymmetry in score distributions.
Thus when using the score, there is no need to estimate the mean of the asset returns to select a portfolio
In the next section, we evaluate the score-based portfolios in a pseudo-real time asset allocation exercise.

\subsection{Asset allocation exercises}

Here, we assess the relevance of the scored-based optimal portfolios in a pseudo-real time asset allocation exercise.
The scored-based optimal portfolios $A$, $B$, $C$ and $D$ are computed in the same way as in Section~\ref{subsec:performance_measures} where, instead of drawing asset returns from a given distribution, we compute the performance measures directly with the in-sample observations. 
By the direct use of the empirical distribution of scores, we avoid to estimate the mean and the covariance matrix of the asset returns. 
We set $S^*$ to $0.5$ to compute the portfolios $A$ and $C$.
These portfolios are compared with the MV optimal portfolio computed using the shrinkage estimate of the covariance matrix of \cite{LW04} and with the Equally-weighted Risk Contributions (ERC) portfolio by \cite{MRT2010} which also uses the shrinkage estimator.
For a fair comparison, all allocations are constrained to have the same volatility which is fixed to the average in-sample volatility of the long-only portfolios. 
This volatility is estimated by sampling portfolios uniformly from $\Delta^{n-1}$, see \cite{RbMel98}, and averaging their volatility.
And, for exhaustiveness, we also compare the unconstrained optimal portfolios. 

Regarding the data, we use two of the French's data sets\footnote{\scriptsize\url{https://mba.tuck.dartmouth.edu/pages/faculty/ken.french/data_library.html}}: the 10 and 30 industry portfolios data sets.
We consider two data sets of different size for robustness check and because a larger number of assets should favor the portfolio using the shrinkage estimator, the MV and ERC portfolios.
The data is monthly and ranges from July 1926 to June 2020.
The allocations are computed over a rolling window of 120 months and their returns are calculated over the following (out-of-sample) month.
Hence, we have 1008 out-of-sample portfolio returns.

\begin{figure}[h!]
\centering
\includegraphics[width=1\textwidth]{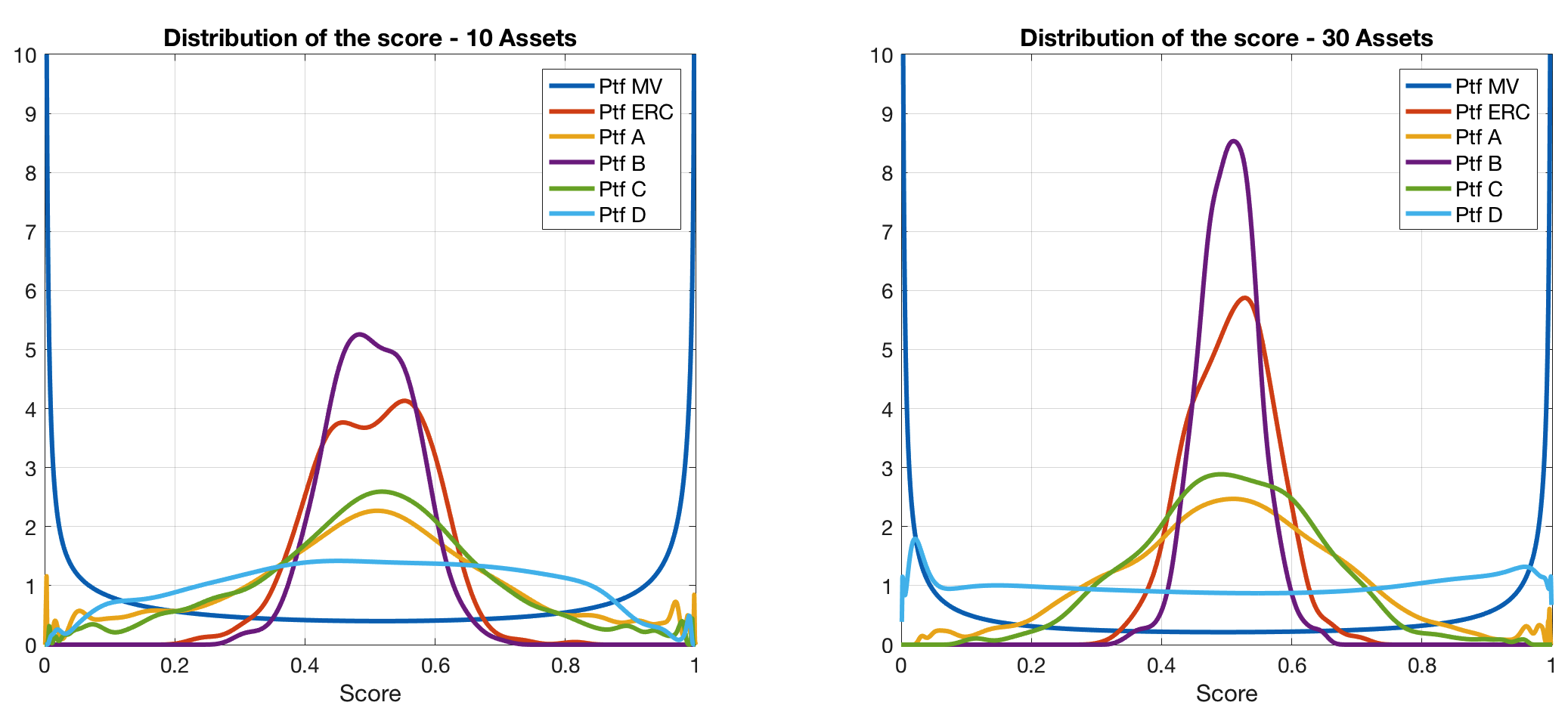}
\caption{Estimates of the optimal portfolios' distributions of scores in the cases of (left) 10 assets and (right) 30 assets, and with constrained volatility.}
\label{fig:distributions_scores_optimals}
\end{figure}

To evaluate the performance of the optimal portfolios, we report statistics of their out-of-sample returns and scores, Sharpe ratio, turn-over and concentration in Table~\ref{Tab:Perf_OptPtfs_statistics_industry}. 
The estimates of the PDF of their scores are reported in Figure~\ref{fig:distributions_scores_optimals}.

\begin{table}[h!]
\footnotesize
\begin{tabular}{l|rr|rrr|rr|r|rr}
 Ptf & \multicolumn{2}{c|}{Returns} & \multicolumn{3}{c|}{Scores} & \multicolumn{2}{c|}{Sharpe ratio}  & Turn & \multicolumn{2}{|c}{Concent.} \\
     & Avg.   & St.D.  & Avg.   & St.D.  & p-val. & IR   & EW Ptf &  over     & Avg. & St.D.    \\
\hline    
\multicolumn{11}{c}{10 assets and constrained volatility} \\
\hline    
 MV  & 0.70\% & 4.61\% & 50.4\% & 39.9\% & 0.74 & 0.15 &  0.01 &  18.0\% & 143\% & 20\% \\
 ERC & 0.69\% & 4.41\% & 50.3\% &  8.6\% & 0.27 & 0.16 &  0.00 &   0.5\% &  11\% &  4\% \\
 A   & 0.69\% & 4.39\% & 50.2\% & 20.7\% & 0.75 & 0.16 & -0.02 &   8.6\% &  51\% & 29\% \\
 B   & 0.69\% & 4.40\% & 50.1\% &  6.7\% & 0.50 & 0.16 & -0.00 &   0.7\% &  10\% &  4\% \\
 C   & 0.69\% & 4.39\% & 50.2\% & 17.9\% & 0.67 & 0.16 & -0.01 &   7.9\% &  42\% & 23\% \\
 D   & 0.69\% & 4.37\% & 50.5\% & 22.6\% & 0.52 & 0.16 & -0.00 &   7.4\% &  58\% & 13\% \\
\hline    
\multicolumn{11}{c}{30 assets and constrained volatility} \\
\hline    
 MV  & 0.73\% & 5.09\% & 49.4\% & 44.2\% & 0.67 & 0.14 &  0.00 &  22.3\% & 172\% & 13\% \\
 ERC & 0.73\% & 4.83\% & 50.7\% & 6.4\% & 0.00 & 0.15 &  -0.00 &  0.4\% &  8\% &  3\% \\
 A   & 0.73\% & 4.79\% & 50.6\% & 17.0\% & 0.23 & 0.15 &  0.00 &  7.0\% & 41\% & 20\% \\
 B   & 0.72\% & 4.81\% & 50.3\% & 4.6\% & 0.04 & 0.15 &  -0.01 &  0.8\% &  7\% &  3\% \\
 C   & 0.73\% & 4.80\% & 50.7\% & 13.4\% & 0.08 & 0.15 &  0.00 &  7.8\% & 31\% & 10\% \\
 D   & 0.73\% & 4.81\% & 50.7\% & 29.8\% & 0.47 & 0.15 &  0.00 &  12.3\% & 90\% & 11\% \\
\hline  
\hline  
\multicolumn{11}{c}{10 assets and unconstrained volatility} \\
\hline  
 MV  & 0.65\% & 4.44\% & 49.8\% & 41.1\% & 0.87 & 0.15 & -0.02 &  16.1\% & 148\% & 24\% \\
 ERC & 0.69\% & 4.12\% & 50.0\% & 14.2\% & 0.96 & 0.17 & 0.00 &  0.5\% & 19\% &  5\% \\
 A   & 0.69\% & 4.32\% & 50.5\% & 16.6\% & 0.32 & 0.16 & 0.01 &  7.1\% & 35\% & 29\% \\
 B   & 0.69\% & 4.32\% & 50.2\% & 2.6\% & 0.01 & 0.16 & 0.01 &  0.2\% &  3\% &  0\% \\
 C   & 0.70\% & 4.30\% & 50.6\% & 11.1\% & 0.10 & 0.16 & 0.03 &  5.4\% & 22\% & 16\% \\
 D   & 0.70\% & 4.33\% & 50.8\% & 23.4\% & 0.25 & 0.16 & 0.02 &  7.2\% & 61\% & 12\% \\
\hline  
\multicolumn{11}{c}{30 assets and unconstrained volatility} \\
\hline  
 MV  & 0.69\% & 5.26\% & 47.7\% & 44.7\% & 0.10 & 0.13 & -0.01 &  23.1\% & 173\% & 12\% \\
 ERC & 0.72\% & 4.53\% & 50.4\% & 18.1\% & 0.43 & 0.16 & -0.00 &  0.6\% & 19\% &  5\% \\
 A   & 0.72\% & 4.76\% & 50.5\% & 10.6\% & 0.16 & 0.15 &  0.01 &  4.3\% & 19\% & 22\% \\
 B   & 0.72\% & 4.77\% & 50.2\% & 1.7\% & 0.00 & 0.15 & 0.00 &  0.3\% &  3\% &  1\% \\
 C   & 0.72\% & 4.76\% & 50.5\% & 6.1\% & 0.02 & 0.15 & 0.01 &  3.7\% & 13\% &  7\% \\
 D   & 0.73\% & 4.84\% & 50.5\% & 30.6\% & 0.60 & 0.15 & 0.02 &  12.6\% & 93\% & 11\% \\
\hline  
\end{tabular}
\caption{Statistics on the performance of the optimal portfolios for the data sets of 10 and 30 Industry asset returns from July 1926 to June 2020. Each Sharpe ratio is computed for a different benchmark: ``EW Ptf" stands for the equally weighted portfolio and ``IR" stands for the Interest Rate. The p-value is for the t-test testing $H_0$: the score of the portfolio is equal to 50\%.}
\label{Tab:Perf_OptPtfs_statistics_industry}
\end{table}

With a constrained volatility and with 10 assets, we find that all optimal portfolios provide similar average returns.
Their out-of-sample volatilities are also very close one another except for the MV portfolio which is a bit more volatile.
In terms of scores, their average scores are not significantly different from 50\%.
The main difference among the optimal portfolios concerns their dispersion. 
Portfolio B and the ERC portfolio show little dispersion in their out-of-sample scores.
This is due to their very low concentration which makes them close to the equally weighted portfolio and implies a very low turn-over.
Portfolios A, C and D present more dispersed scores, which can be related to their concentration and turn-over.
Finally the MV portfolio has by far the most dispersed scores, the highest concentration and the highest turn-over
The same observations are made with the 30 assets, with the average score of portfolios B and ERC being now significantly different from 50\%.

When the optimal portfolios' volatility is unconstrained, the returns of the MV portfolio deteriorate comparatively to the other optimal portfolios with a lower average return and an higher volatility.
On the contrary, the ERC portfolio performs better than its score-based opponents with a lower volatility but a similar average return.
In term of scores, the observations made for the case of constrained volatility hold, with the average score of portfolios B being significantly different from 50\% in the two data sets and the average score of portfolios C in the case of 30 assets.  
The ERC portfolio also tends to be more concentrated than portfolio B, with still very low turn-over for both portfolios.  

Overall, the MV portfolio performs worse than the ERC and the score-based portfolios in all cases.
Whereas the ERC and the score-based portfolios appear to perform quite similarly, with a slight advantage to the ERC portfolio when the optimal portfolio volatility in unconstrained.

\section{Concluding remarks and future work}\label{Sconcl}
								
In this paper, we reviewed various approaches to compute the PDF, CDF and moments of the distribution of portfolio returns, across portfolios and for the long-only strategy.

For the CDF, the computations improve upon existing work by providing exact results, allowing for equal asset returns, and handle a large number of assets, thus removing the need of Monte Carlo sampling for its estimation.
These computations can be based on:
\begin{itemize}
	\item the volume algorithm by~\cite{Varsi} which is fast and exact even for a large number of assets,
	\item the closed form expression of CDF, which is exact but numerically unstable for a large number of assets ($>20$).
\end{itemize}
For the PDF, our methods are new and based on one of the following three approaches:
\begin{itemize}
	\item the algorithm by de Boor and Cox \cite{dB72,C72}, which is exact but too slow for a large number (namely $>20$) of assets to be of practical use,
	\item the closed form expression of PDF, which is exact but numerically unstable for a large number ($>20$) of assets,
	\item the numerical derivation of the CDF using \cite{Varsi}, which only provides an estimate but is fast and applies to a large number of assets.
\end{itemize}	
For the moments, the computations are new and can based on
\begin{itemize}
	\item closed form expressions up to the fourth-order moment,
	\item a new algorithm using Algorithm $ZS1$ by \cite{ZS1994}, for higher moments, which is fast and exact even for a large number of assets.
\end{itemize}
It should be noted that most of these computations can easily be vectorized, thus further extending their realm of applications. \\

These results have several statistical implications in financial modeling. 
\begin{itemize}
	\item The asset returns can be standardized cross-sectionally without altering the relative performance of portfolios. The series obtained are then robust to systemic heteroskedasticity.
	\item The closed form expressions of the first four moments show a direct mapping between the moments of the cross-sectional asset returns distribution and those of the distribution of portfolio returns.
	
	In particular, the first moments are identical for both distributions. The second and third moments are proportional to each other, the factor depending on the number of assets.
	For the second moment, the factor is $\frac{1}{N+1}$ where $N$ is the number of assets, implying that it might be more difficult to find a portfolio which performs significantly better than the equally weighted portfolio, when the number of assets increases.
	
	The closed form expression of the fourth moment behaves differently with an extra positive term usually implying fatter tails in the distribution of portfolio returns than in the cross-sectional distribution of the asset returns.
	Its implications have to be further analyzed.
\end{itemize}

One remark is due on the relevance of computing high moments.
We believe that these moments should be useful in an alternative method recovering the PDF.
Indeed, since the distribution is bounded, this problem is known as the Hausdorff moment problem, which has been addressed, see for instance \cite{MNATSAKANOV20081869} and \cite{BJL2019}.
This approach has been inconclusive for the authors.

Regarding the applications, we illustrated the relevance of the score in asset allocation.
In particular, we have seen that
\begin{itemize}
\item the distribution of score offers a new way to compare portfolios, 
\item by introducing new portfolio performance measures based on the score, we can get optimal portfolios which are less concentrated, less risky in terms of score and with better average score than the MV portfolio,
\item the score-based optimal portfolios outperform the MV portfolio and compare with the ERC portfolio.
\end{itemize}



 
Future work can take several directions. 
On the theoretical side, it would be interesting to study the distribution of portfolio volatilities in order to better assess the dependency between portfolios returns and volatilities, which is done by sampling in \cite{CCEF2018}.
It would also be interesting to consider different investment sets, e.g.\ including short selling (i.e.\ negative portfolio weights) and leverage (i.e.\ sum of portfolio weights larger than one).
In terms of applications, it should be noticed that the paper focuses on portfolio returns but the methodology can be applied to any linear combination of asset characteristics, e.g. by defining the portfolio dividend yield as the weighted sum of the asset dividend yields.
Also, the score robustness to even shocks and to volatility jumps would make it interesting to study the persistence of the shape of the score distributions.
Also the relationship between concentration and the shape of the score distributions has to be further studied.

\clearpage

\bibliographystyle{apalike}
\bibliography{Biblio_PoC_APS}

\clearpage

\appendix

\section{Proof of Lemma~\ref{Lemma_id_2}}\label{Ax5}

{\bf Lemma~\ref{Lemma_id_2}.}
For $n \in \NN$, it holds 
\[
6 \sum_{i=1}^n \sum_{j=i}^n \sum_{k=j}^n x_i x_j x_k = \left( \sum_{i=1}^n x_i \right)^3 + 3 \left( \sum_{i=1}^n x_i \right) \left( \sum_{i=1}^n x_i^2 \right) + 2 \sum_{i=1}^n x_i^3 .
\]

\noindent{\it Proof}. 
Let $S(n) = \sum_{i=1}^n \sum_{j=i}^n \sum_{k=j}^n x_i x_j x_k$. The cases $n=0$ and $n=1$ are easily verifiable. Let us assume that the theorem holds for $S(n), n \geq 1$. We shall prove it for $S(n + 1)$. Clearly,

\[  S(n+1) = S(n) + \sum_{i=1}^n \sum_{j=i}^n x_i x_j x_{n+1} + \sum_{i=1}^n x_i x_{n+1}^2 + x_{n+1}^3, 
\]
which, by inductive hypothesis, yields 
\begin{align*}
& 6 S(n + 1) = \\
& \left( \sum_{i=1}^n x_i \right)^3 + 3 \left( \sum_{i=1}^n x_i \right) \left( \sum_{i=1}^n x_i^2 \right) + 2 \sum_{i=1}^n x_i^3 + \underline{6 x_{n+1} \sum_{i=1}^n \sum_{j=i}^n x_i x_j} + 6 x_{n+1}^2 \sum_{i=1}^n x_i + 6 x_{n+1}^3 .
\end{align*}
The underlined term becomes 
\begin{equation}\label{eq_sum}
3 x_{n+1} \left( 2 \sum_{i=1}^n x_i^2 + 2 \sum_{i=1}^n \sum_{j>i}^n x_i x_j\right),
\end{equation}
where the last index $j$ is strictly larger than $i$. The overall sum is re-written as the sum of the following three terms, where we have underlined terms corresponding to sum (\ref{eq_sum}):
$$
\left( \sum_{i=1}^n x_i \right)^3 + \underline{3 x_{n+1} \left( \sum_{i=1}^n x_i \right)^2} + 3 x_{n+1}^2 \sum_{i=1}^n x_i + x_{n+1}^3 = \left( \sum_{i=1}^n x_i + x_{n+1} \right)^3,
$$
\begin{align*}
	& 3 \left[ \left( \sum_{i=1}^n x_i \right) \left( \sum_{i=1}^n x_i^2 \right) + \underline{x_{n+1} \sum_{i=1}^n x_i^2}  + x_{n+1}^2 \sum_{i=1}^n x_i + x_{n+1}^3 \right] =\\
= & 3 \left(  \sum_{i=1}^n x_i + x_{n+1}\right)\left( \sum_{i=1}^n x_i^2 + x_{n+1}^2 \right),
\end{align*}
\[
2 \sum_{i=1}^n x_i^3 + 2 x_{n+1}^3,
\]
which correspond to the three sums of the original claim.


\section{Proof of Lemma~\ref{Lemma_id_3}}\label{Ax6}

{\bf Lemma \ref{Lemma_id_3}.}
For $n \in \NN$, it holds 
\begin{align*}
  & (4!) \sum_{i=1}^n \sum_{j=i}^n \sum_{k=j}^n \sum_{t=k}^n x_i x_j x_k x_t =\\
& \left( \sum_{i=1}^n x_i \right)^4 + 6 \left( \sum_{i=1}^n x_i \right)^2 \left( \sum_{i=1}^n x_i^2 \right) + 8 \left( \sum_{i=1}^n x_i \right) \left( \sum_{i=1}^n x_i^3 \right) + 6 \left( \sum_{i=1}^n x_i^4 \right) + 3 \left( \sum_{i=1}^n x_i^2 \right)^2\\
\end{align*}

\noindent{\em Proof}. 
Let $S(n) = \sum_{i=1}^n \sum_{j=i}^n \sum_{k=j}^n \sum_{t=k}^n x_i x_j x_k x_t$. The cases $n=0$ and $n=1$ are easily verifiable. Let us assume that the theorem holds for $S(n), n \geq 1$. We shall prove it for $S(n + 1)$. Clearly,

\[  S(n+1) = S(n) + \sum_{i=1}^n \sum_{j=i}^n \sum_{k=j}^n x_i x_j x_k x_{n+1} + \sum_{i=1}^n \sum_{j=i}^n x_i x_j x^2_{n+1} +  \sum_{i=1}^n x_i x_{n+1}^3 + x_{n+1}^4, \]

which, by the inductive hypothesis, yields: 
\begin{align*}
		& 24 S(n + 1) =\\
& \left( \sum_{i=1}^n x_i \right)^4 + 6 \left( \sum_{i=1}^n x_i \right)^2 \left( \sum_{i=1}^n x_i^2 \right) + 8 \left( \sum_{i=1}^n x_i \right) \left( \sum_{i=1}^n x_i^3 \right) + 6 \left( \sum_{i=1}^n x_i^4 \right) + 3 \left( \sum_{i=1}^n x_i^2 \right)^2\\
		& + 24 x_{n+1} \sum_{i=1}^n \sum_{j=i}^n \sum_{k=j}^n x_i x_j x_k  + 24 x^2_{n+1} \sum_{i=1}^n \sum_{j=i}^n x_i x_j + 24 x_{n+1}^3 \sum_{i=1}^n x_i + 24 x_{n+1}^4.
\end{align*}
Let $ A = x_{n+1}^3 \sum_{i=1}^n x_i$ and $B = x_{n+1}^4$.
From Lemma \ref{Lemma_id_2}, we have
\begin{align*}
		& 24 x_{n+1} \sum_{i=1}^n \sum_{j=i}^n \sum_{k=j}^n x_i x_j x_k =\\
= 	& 4\, x_{n+1} \left( \left( \sum_{i=1}^n x_i \right)^3 + 3 \left( \sum_{i=1}^n x_i \right) \left( \sum_{i=1}^n x_i^2 \right) + 2 \sum_{i=1}^n x_i^3 \right)\\
= 	& \underbrace{4 x_{n+1} \left( \sum_{i=1}^n x_i \right)^3}_{ = C} + \underbrace{12 x_{n+1} \left( \sum_{i=1}^n x_i \right) \left( \sum_{i=1}^n x_i^2 \right)}_{ = D} + \underbrace{8 x_{n+1} \sum_{i=1}^n x_i^3}_{ = E}
\end{align*}

Then, from Lemma~\ref{Lemma_id_2}, we have
\begin{align*}
	& 24 x^2_{n+1} \sum_{i=1}^n \sum_{j=i}^n x_i x_j =
	12\, x^2_{n+1} \left( \left( \sum_{i=1}^n x_i \right)^2 + \sum_{i=1}^n x_i^2 \right) =\\
= 	& \underbrace{12\, x^2_{n+1} \left( \sum_{i=1}^n x_i \right)^2}_{ = 2 F} + \underbrace{12 x^2_{n+1} \left( \sum_{i=1}^n x_i^2 \right)}_{ = 2 G} .
\end{align*}
The overall sum is rewritten using the sum of the following four terms:
\begin{align*}
\left( \sum_{i=1}^{n+1} x_i \right)^4 = & \left( \sum_{i=1}^n x_i \right)^4 + \underbrace{4 x_{n+1} \left( \sum_{i=1}^n x_i \right)^3}_{=C} + \underbrace{6 x_{n+1}^2 \left( \sum_{i=1}^n x_i \right)^2}_{=F} + \underbrace{4 x_{n+1}^3 \left( \sum_{i=1}^n x_i \right)}_{=4A} + \underbrace{x_{n+1}^4}_{=B} .
\end{align*}

\begin{align*}
6 \left( \left( \sum_{i=1}^{n+1} x_i \right)^2 \left( \sum_{i=1}^{n+1} x_i^2 \right) \right) = &  6 \left( \left( \left( \sum_{i=1}^n x_i \right)^2 + 2 x_{n+1} \left( \sum_{i=1}^n x_i \right) + x^2_{n+1} \right) \left( x^2_{n+1} + \sum_{i=1}^n x_i^2 \right) \right) \\
  = & 6 \left( \sum_{i=1}^n x_i \right)^2 \left( \sum_{i=1}^n x_i^2 \right) + \underbrace{ 6 x^2_{n+1} \left( \sum_{i=1}^n x_i \right)^2}_{=F} + \underbrace{ 12 x^3_{n+1} \left( \sum_{i=1}^n x_i \right)}_{=12A} \\
	  & + \underbrace{ 12  x_{n+1} \left( \sum_{i=1}^n x_i \right) \left( \sum_{i=1}^n x_i^2 \right)}_{=D} + \underbrace{ 6 x^4_{n+1}}_{=6B} + \underbrace{x^2_{n+1} \left( \sum_{i=1}^n x_i^2 \right)}_{=G} .
\end{align*}

\begin{align*}
8 \left( \sum_{i=1}^{n+1} x_i \right) \left( \sum_{i=1}^{n+1} x_i^3 \right) = & 8 \left( \sum_{i=1}^{n} x_i \right) \left( \sum_{i=1}^{n} x_i^3 \right) + \underbrace{ 8 x_{n+1} \left( \sum_{i=1}^n x_i^3 \right)}_{=E} + \underbrace{ 8 x_{n+1}^3 \left( \sum_{i=1}^n x_i \right)}_{=8A}  + \underbrace{ 8 x_{n+1}^4}_{=8B} .
\end{align*}

\begin{align*}
3 \left( \sum_{i=1}^{n+1} x_i^2 \right)^2 = & 3 \left( \sum_{i=1}^n x_i^2 \right)^2 + \underbrace{6 x^2_{n+1} \left( \sum_{i=1}^n x_i^2 \right)}_{=G} + \underbrace{3 x_{n+1}^4}_{=3B} .
\end{align*}

\begin{align*}
6 \left( \sum_{i=1}^{n+1} x_i^4 \right)  = 6 \left( \sum_{i=1}^n x_i^4 \right) + \underbrace{6 x_{n+1}^4}_{=6B} .
\end{align*}

\section{Appendix to the applications}\label{appendix:data}

\subsection{Estimated variance-covariance matrix}\label{subSec:EstCovMatrix}

The estimation of the covariance matrix $\Sigma$ is: 

\[
\tiny
\Sigma = 10^{-3} \left[ \begin{array}{rrrrrrrrrr} 
    1.3101  &  1.6200  &  1.2866  &  1.0601  &  1.2428  &  1.1467  &  1.0406  &  0.8504  &  0.9328  &  1.3467\\
    1.6200  &  6.9273  &  3.3182  &  2.2277  &  4.0887  &  3.1160  &  2.7040  &  1.4480  &  1.7163  &  3.3322\\
    1.2866  &  3.3182  &  2.9002  &  1.9157  &  2.8463  &  2.0617  &  1.8071  &  1.1666  &  1.3364  &  2.2969\\
    1.0601  &  2.2277  &  1.9157  &  3.5962  &  2.0260  &  1.5902  &  1.1932  &  0.9238  &  1.7941  &  1.6636\\
    1.2428  &  4.0887  &  2.8463  &  2.0260  &  7.2544  &  3.4768  &  2.4313  &  1.5724  &  1.2506  &  2.7809\\
    1.1467  &  3.1160  &  2.0617  &  1.5902  &  3.4768  &  3.6761  &  1.8619  &  1.2328  &  1.0867  &  2.1781\\
    1.0406  &  2.7040  &  1.8071  &  1.1932  &  2.4313  &  1.8619  &  2.2896  &  0.8535  &  0.9041  &  1.9379\\
    0.8504  &  1.4480  &  1.1666  &  0.9238  &  1.5724  &  1.2328  &  0.8535  &  1.6401  &  0.9468  &  1.2521\\
    0.9328  &  1.7163  &  1.3364  &  1.7941  &  1.2506  &  1.0867  &  0.9041  &  0.9468  &  2.2781  &  1.3937\\
    1.3467  &  3.3322  &  2.2969  &  1.6636  &  2.7809  &  2.1781  &  1.9379  &  1.2521  &  1.3937  &  3.2168\\
\end{array} \right] 
\]

\subsection{Table of differences among the 4 portfolios}\label{subsec:dif_ptfs}

The four portfolios differ in two aspects: (1) they are distant one another (2) their concentrations differ, with the first portfolio being the most concentrated (119\%) and the fourth portfolio being the least concentrated (35\%).

\begin{table}[h!]
\centering
\begin{tabular}{l|ccc|c}
	  & Ptf 2  & Ptf 3 	& Ptf 4 & EWPtf\\
\hline
Ptf 1 & 82\%   &   95\% &  133\%   &  119\%\\
Ptf 2 &        &   59\% &  74\%    &  61\%\\
Ptf 3 &        &        &  59\%    &  49\%\\
Ptf 4 &        &        &          &  35\%
\end{tabular}
\caption{Turn-over distance between the portfolios; EWPtf stands for the equally weighted portfolio.}
\label{tab:TO_4Ptfs}
\end{table}

\subsection{Skew-t distributions' parameters}\label{subSec:SkewTParameters}

Let denote $\mathbf{0}$ and $\mathbf{1}$ the $10 \times 1$ vectors of 0's and of 1's, respectively. $\mathbf{I}$ denotes the identity matrix.
The parameters of the asset returns' skew-t distribution are reported according to the Centred Parameterization, see \cite{AA2013} for details.. 
The mean is $\mathbf{0}$ and the variance-covariance matrix $0.0035 \ \mathbf{I}$. 
Our three cases set different skewness coefficients $\gamma_1$ and measure of kurtosis $\gamma_2$: no skewness ($\gamma_1=\mathbf{0}$ and $\gamma_2=100$), the same skewness ($\gamma_1=-0.3 \ \mathbf{1}$ and $\gamma_2=100$) and different skewnesses ($\gamma_1=(0, 0, 0, -0.7, 0, 0, 0, 0, -0.7, 0)$ and $\gamma_2=100$).

We use the random number generation for the multivariate skew-t of the R package \textit{sn} by Azzalini.\footnote{https://cran.r-project.org/package=sn}

\subsection{Generalized hyperbolic distributions' parameters}\label{subSec:GHypParameters}

The parameters of the Generalized hyperbolic parameters have been computed using the 10 industry indices of French's dataset ranging from January 2000 to December 2009.
We do not aim for the best estimates but for realistic values of parameters.
The mean and covariance matrix of the Generalized hyperbolic distribution have been set to the sample mean and to $\Sigma$, respectively.
The rest of the parameters of the Generalized hyperbolic parameters have been fitted on the data using the R package \textit{ghyp}\footnote{https://cran.r-project.org/package=ghyp} and their values are $\lambda = 1.3845$, $\bar{\alpha} = 1.8082$ and
\[ \gamma = - \left[ \begin{array}{r}  0.0071 \\ 0.0113 \\ 0.0159 \\ 0.0086 \\ 0.0207 \\ 0.0220 \\ 0.0152 \\ 0.0042 \\ 0.0095 \\ 0.0135 \end{array} \right] \]

\clearpage

\subsection{Estimates of the distributions of the optimal portfolios' score - Gaussian case}\label{subSec:ScoreDist_Gaussian}

\begin{figure}[h!]
\centering
\includegraphics[width=0.8\textwidth]{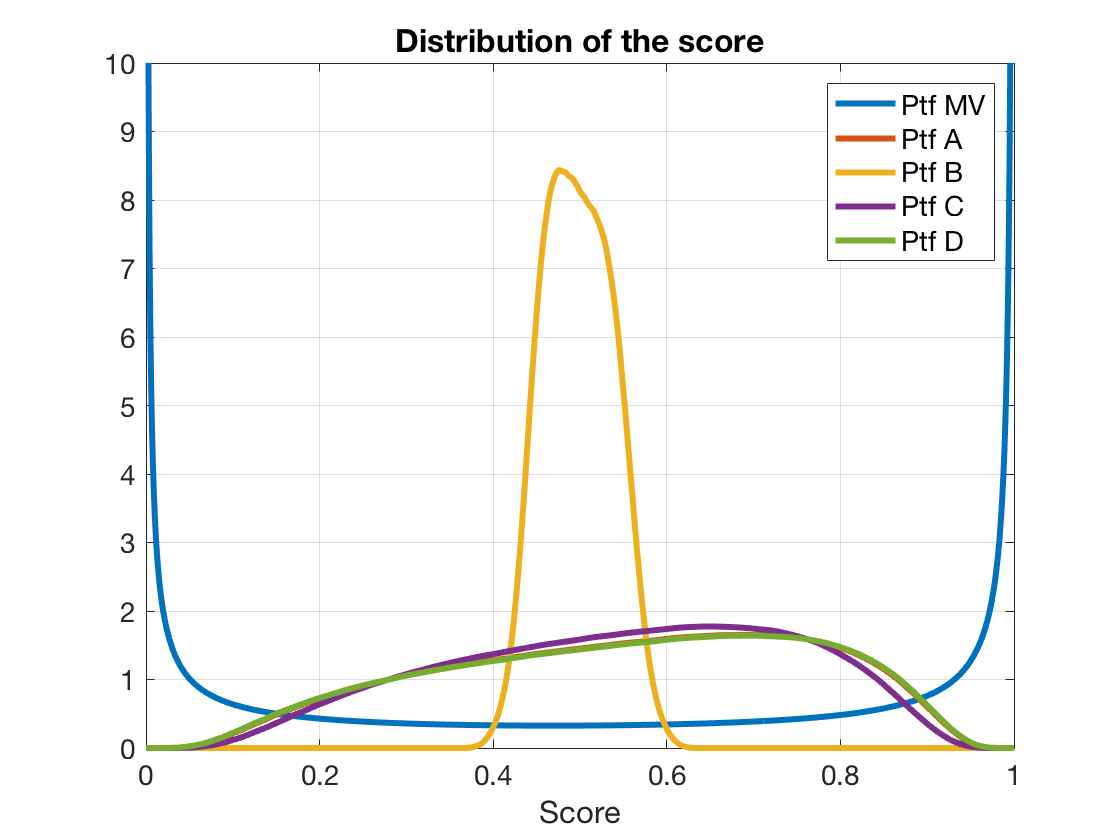}
\caption{Estimates of the distributions of the scores of the optimal portfolios when the asset returns follow the Gaussian distribution.}
\label{Fig:Score_MVSc}
\end{figure}

\subsection{Probabilities on the optimal portfolios' scores - Gaussian case}\label{subSec:ProbaScores_Gaussian}

\begin{table}[h!]
\centering
\begin{tabular}{l|cccc}
			& \multicolumn{4}{|c}{Probability of a score being}\\
			& between $20\%$ and $80\%$ & above $50\%$ &  below $10\%$ & above $90\%$\\
\hline						
MV-Opt.	&   $21.9\%$	&   $57.6\%$	&   $26.6\%$	&   $40.4\%$	\\
$Perf_A$	&   $81.9\%$	&   $60.2\%$	&    $0.5\%$	&    $1.6\%$	\\
$Perf_B$	&  	$100\%$		&   $47.1\%$	&       $0\%$	&       $0\%$	\\
$Perf_C$	&   $86.3\%$	&   $60.2\%$	&    $0.2\%$	&    $0.8\%$	\\
$Perf_D$	&   $81.6\%$	&   $60.2\%$	&    $0.6\%$	&    $1.7\%$	\\
\end{tabular}
\caption{Empirical probability to observe a score being between $20\%$ and $80\%$, being above $50\%$, being below $10\%$, and being above $90\%$, for each optimal portfolio when the asset returns follow the multivariate Gaussian distribution.}
\label{Tab:Probas_Opt_C}
\end{table}

\subsection{Distances among the optimal portfolios - Gaussian case}\label{subSec:PtfsDistances_Gaussian}

\begin{table}[h!]
\centering
\begin{tabular}{l|ccccc}
	  & $Perf_A$ & $Perf_B$  & $Perf_C$ 	& $Perf_D$ & EWPtf\\
\hline
Mean-Var & 115\% & 142\% & 118\% & 114\% & 140\% \\
$Perf_A$ & & 44.2\% & 5.5\% & 0.62\% & 44.4\% \\
$Perf_B$ & & & 39.7\% & 44.4\% & 4.2\% \\
$Perf_C$ & & & & 5.9\% & 39.6\% \\
$Perf_D$ & & & & & 44.6\% \\
\end{tabular}
\caption{Turn-over distance between the optimal portfolios when the asset returns follow the multivariate Gaussian distribution; EWPtf stands for the equally weighted portfolio.}
\label{tab:TO_Opt_Ptfs_gaussian}
\end{table}

\subsection{Efficient frontier - Gaussian case}\label{subSec:EfficientFrontier_Gaussian}

\begin{figure}[h!]
\centering
\includegraphics[width=1.0\textwidth]{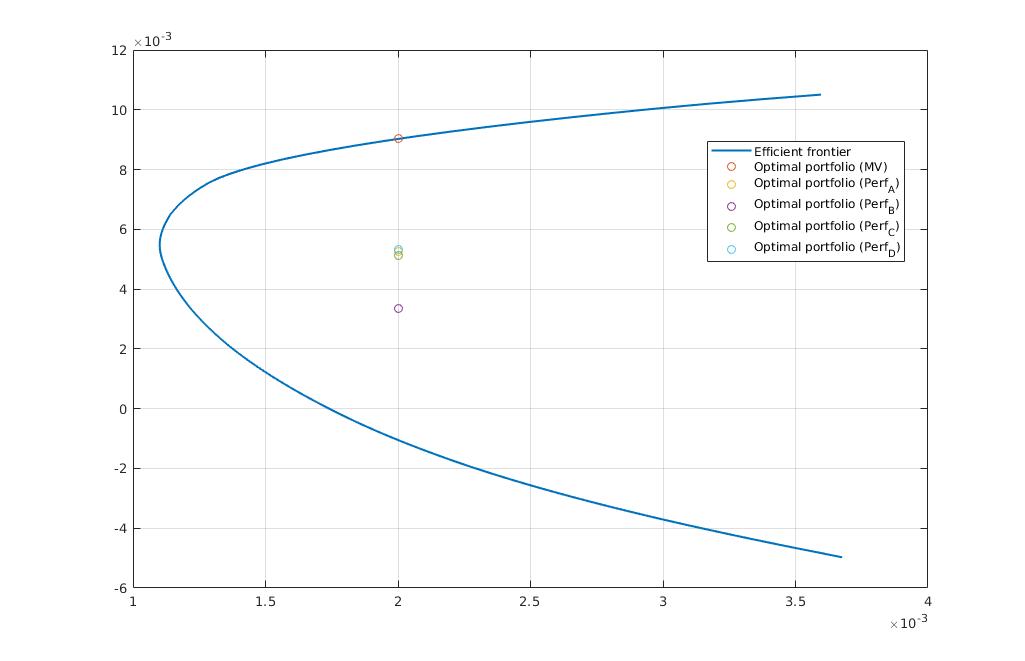}
\caption{Efficient frontier with the optimal portfolios highlighted when the asset returns follow the multivariate Gaussian distribution.}
\label{Fig:EfficientFrontier}
\end{figure}

\subsection{Estimates of the distributions of the optimal portfolios' score - Generalized Hyperbolic case}\label{subSec:ScoreDist_GHyp}

\begin{figure}[h!]
\centering
\includegraphics[width=0.75\textwidth]{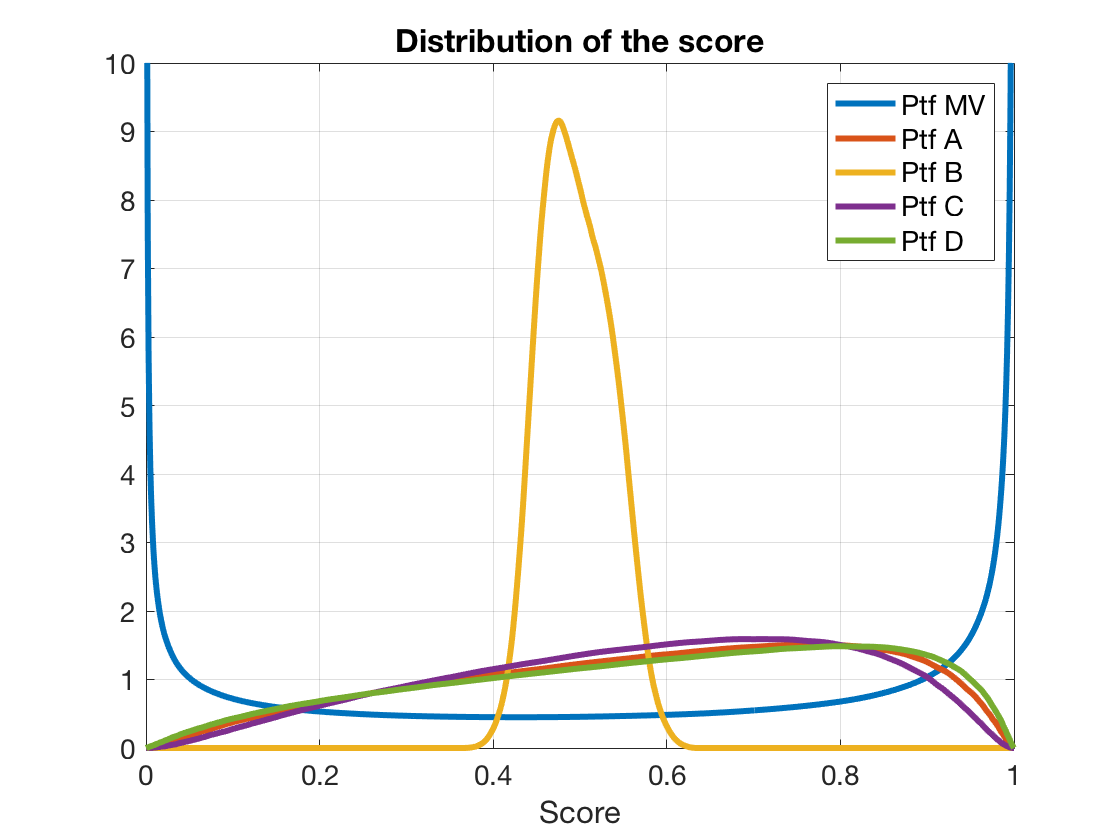}
\caption{Estimates of the distributions of the scores of the optimal portfolios when the asset returns follow the Multivariate  Generalized Hyperbolic distribution.}
\label{Fig:Score_MVSc_ghyp}
\end{figure}

\subsection{Probabilities on the optimal portfolios' scores - Generalized Hyperbolic case}\label{subSec:ProbaScores_GHyp}

\begin{table}[h!]
\centering
\begin{tabular}{l|cccc}
			& \multicolumn{4}{|c}{Probability of a score being}\\
			& between $20\%$ and $80\%$ & above $50\%$ &  below $10\%$ & above $90\%$\\
\hline						
MV-Opt.	&   $29.9\%$	&   $60.8\%$	&   $19.0\%$	&   $36.7\%$	\\
$Perf_A$	&   $71.3\%$	&   $63.9\%$	&    $1.8\%$	&    $7.4\%$	\\
$Perf_B$	&  	$100\%$		&   $44.1\%$	&       $0\%$	&       $0\%$	\\
$Perf_C$	&   $76.3\%$	&   $63.8\%$	&    $1.1\%$	&    $4.9\%$	\\
$Perf_D$	&   $68.6\%$	&   $63.9\%$	&    $2.3\%$	&    $8.9\%$	\\
\end{tabular}
\caption{Empirical probability to observe a score being between $20\%$ and $80\%$, being above $50\%$, being below $10\%$, and being above $90\%$, for each optimal portfolio when the asset returns follow the multivariate Generalized Hyperbolic distribution.}
\label{Tab:Probas_Opt_C_ghyp}
\end{table}

%

\subsection{Distances among the optimal portfolios - Generalized Hyperbolic case}\label{subSec:PtfsDistances_GHyp}

\begin{table}[h!]
\centering
\begin{tabular}{l|ccccc}
	  & $Perf_A$ & $Perf_B$  & $Perf_C$ 	& $Perf_D$ & EWPtf\\
\hline
Mean-Var & 1.14\% & 160\% & 120\% & 111\% & 160\% \\
$Perf_A$ &  & 72.4\% & 7.3\% & 4.2\% & 72.1\% \\
$Perf_B$ &  &  & 65.9\% & 75.9\% & 4.1\% \\
$Perf_C$ &  &  &  & 11.5\% & 65.6\% \\
$Perf_D$ &  &  &  &  & 75.6\% \\
\end{tabular}
\caption{Turn-over distance between the optimal portfolios when the asset returns follow the multivariate Generalized Hyperbolic distribution; EWPtf stands for the equally weighted portfolio.}
\label{tab:TO_Opt_Ptfs_ghyp}
\end{table}

\end{document}